\documentclass[%
reprint,
superscriptaddress,
amsmath,amssymb,
aps,
prl
]{revtex4-1}
\usepackage{graphicx}
\usepackage{dcolumn}
\usepackage{bm}
\usepackage{physics}
\usepackage{color}
\RequirePackage[colorlinks=true,linkcolor=blue,urlcolor=blue,hyperfootnotes=false,citecolor=black]{hyperref}

\begin{document}
\title{Synthesizing multi-phonon quantum superposition states using flux-mediated three-body interactions with superconducting qubits}

\author{Marios Kounalakis}
 \email{marios.kounalakis@gmail.com}
\author{Yaroslav M. Blanter}
\author{Gary A. Steele}
\affiliation{Kavli Institute of Nanoscience, Delft University of Technology, 2628 CJ Delft, The Netherlands}

\date{\today}

\begin{abstract}
Massive mechanical resonators operating at the quantum scale can enable a large variety of applications in quantum technologies, as well as fundamental tests of quantum theory.
Of crucial importance in that direction, is both their integrability into state-of-the-art quantum platforms as well as the ability to prepare them in generic quantum states using well-controlled high-fidelity operations.
Here, we propose a scheme for controlling a radio-frequency mechanical resonator at the quantum scale using two superconducting transmon qubits that can be integrated on the same chip.
Specifically, we consider two qubits coupled via a capacitor in parallel to a superconducting quantum interference device (SQUID), which has a suspended mechanical beam embedded in one of its arms.
Following a theoretical analysis of the quantum system, we find that this configuration, in combination with an in-plane magnetic field, can give rise to a tuneable three-body interaction in the single-photon strong-coupling regime, while enabling suppression of the stray qubit-qubit coupling.
Using state-of-the-art parameters and qubit operations at single-excitation levels, we numerically demonstrate the possibility of ground-state cooling as well as high-fidelity preparation of mechanical quantum states and qubit-phonon entanglement, i.e. states having negative Wigner functions and obeying non-classical correlations.
Our work significantly extends the quantum control toolbox of radio-frequency mechanical resonators and may serve as a promising architecture for integrating such mechanical elements with transmon-based quantum processors.
\end{abstract}

\maketitle

\section{Introduction}
The ability to control massive mechanical objects at the quantum level constitutes a very interesting task for many technological applications, ranging from microwave-to-optical conversion to quantum memories, as well as fundamental studies regarding the quantum-classical divide~\cite{penrose1996gravity,marshall2003towards, kleckner2008creating, blencowe2011light, blencowe2007quantum, stannigel2012optomechanical, bochmann2013nanomechanical, andrews2014bidirectional, metcalfe2014applications, aspelmeyer2014cavity}.
The rapid development of cavity optomechanics over the last decade has enabled the exploration of mechanical resonators in regimes where quantum effects become prominent.
One approach relies on resonant coupling of acoustic phonons to microwave excitations via piezoelectric materials~\cite{oconnell2010quantum, chu2018creation}, however, the amplitude of the lattice vibrations in these systems is very small, limiting their applications.
On the other hand, in typical opto- and electromechanical setups, low-frequency mechanical resonators are controlled via parametric coupling to an optical or microwave cavity~\cite{chan2011laser, teufel2011sideband, wollman2015quantum, riedinger2018remote, ockeloen2018stabilized}.
The mechanical elements in these systems are usually realised with metal drumheads or beams, which are characterised by large quality factors and large mass, making them also prime candidates for experimental tests of gravity-induced wavefunction collapse theories~\cite{penrose1996gravity, marshall2003towards, kleckner2008creating,  blencowe2011light}.

Following the pioneering work on acoustic resonators~\cite{oconnell2010quantum}, recently, quantum superpositions of the ground and first excited state were for the first time generated in a parametrically coupled mechanical resonator~\cite{reed2017faithful}.
In this approach, a superconducting qubit is used to create the excitation, which decays into a microwave resonator on a different chip and subsequently transferred to the mechanical element via an effective linearised interaction.
It is recognised, however, that this method has severe limitations as it relies on strong driving, which is challenging with qubits, and suffers from unavoidable losses during the state transfer between different chips, limiting the fidelity of the prepared state~\cite{viennot2018phonon}.
A different scheme, implemented in the optical domain, uses entanglement and post-selective measurements to generate single-photon states~\cite{Hong2017handbury}, although the non-deterministic nature of the protocol in combination with low count rates limits the types of states that can be prepared.

A promising route towards high-fidelity mechanical quantum control is the ability to operate in the single-photon strong-coupling regime, where interaction times are faster than dissipation processes, which however still remains an experimental challenge for far-detuned parametrically coupled mechanical resonators~\cite{pirkkalainen2015cavity, viennot2018phonon}.
Operating in this regime is predicted to give rise to non-classical photon correlations~\cite{rabl2011photon} and non-Gaussian states~\cite{nunnenkamp2011single, nation2013nonclassical}, as well as macroscopic mechanical superpositions~\cite{liao2016macroscopic}.
Moreover, using the radiation-pressure coupling to qubits can lead to the creation of mechanical Schr{\"o}dinger cat states~\cite{armour2002entanglement, mehdi2015quantum}.
Generating Fock states in this regime could also be enabled using an additional microwave resonator to create an effective tripartite coupling, as predicted in Ref.~\cite{mehdi2015quantum}, although this approach is limited to low state preparation fidelities mainly due to the limitations of capacitive coupling and stray qubit-resonator interaction.
In spite of the experimental and theoretical advances in the field, high-fidelity quantum state preparation of mechanical systems appears to be limited to a small class of engineerable states which are, to a large degree, architecture-dependent.

Here, we analyse a new scheme for synthesizing generic mechanical states by employing tuneable three-body interactions between two superconducting qubits and a mechanical resonator in the single-photon strong-coupling regime.
The coupling relies on embedding a suspended micrometer-long beam in one of the arms of a superconducting quantum interference device (SQUID) in combination with an externally applied magnetic field~\cite{nation2008quantum, nation2016ultrastrong, shevchuk2017strong, rodrigues2019coupling}.
We find that, by connecting two superconducting transmon qubits~\cite{koch2007charge} directly via this mechanical SQUID, a tuneable three-body interaction arises as the qubit-qubit flux-mediated coupling is modulated by the mechanical displacement.
Importantly, the detrimental exchange-type interaction between the two qubits can be suppressed by adding a capacitor in parallel to the SQUID, as realised in Ref.~\cite{kounalakis2018tuneable}.
Using state-of-the-art parameters, reported in recent experiments~\cite{rodrigues2019coupling}, we numerically demonstrate the possibility of high-fidelity coherent quantum control of the beam, from ground-state cooling to fast and high-fidelity preparation of mechanical Fock states, as well as maximally entangled states of qubits and phonons.
Finally, we devise a protocol consisting of qubit flux-pulsing and post-selective measurements for synthesizing multi-phonon superposition states, extending the quantum control toolbox and the plurality of engineerable quantum states in radio-frequency mechanical resonators.

\section{Results}
\subsection{Motion-dependent qubit-qubit interaction}
\begin{figure}[t]
  \begin{center}
  \includegraphics[width=1\linewidth]{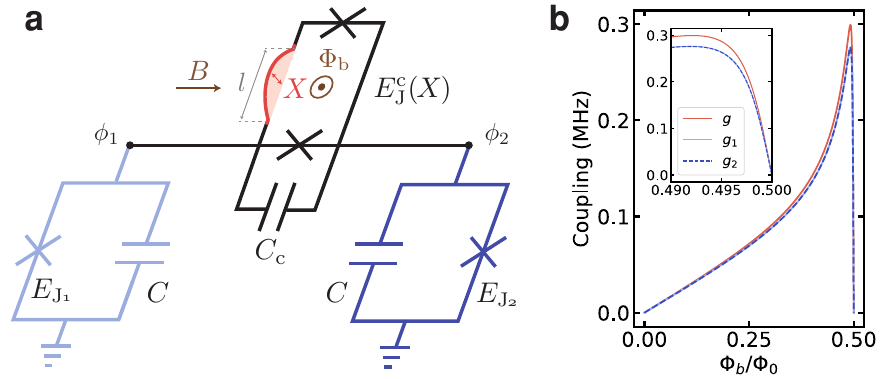}
  \end{center}
  \caption{ {\bf Tripartite coupling architecture.}
  {\bf a.} Circuit diagram of the electromechanical system comprising two transmon qubits directly coupled via a SQUID coupler with an embedded beam that can oscillate out of plane.
  Tuning the coupler to its filter frequency, where linear currents through the capacitor and the SQUID cancel out, and applying an in-plane magnetic field $B$, results in a dominant tripartite coupling as the beam oscillations modulate the qubit-qubit interaction.
  {\bf b.} Flux-mediated couplings as a function of flux bias $\Phi_\text{b}$ for in-plane magnetic field $B=10$~mT and circuit parameters used in this work.
  The red curve represents the tripartite coupling strength, while the solid/dashed blue curves correspond to the radiation-pressure coupling of the beam with each qubit.}
  \label{fig:circuit}
\end{figure}

The proposed circuit, shown in Fig.~\ref{fig:circuit}a, comprises two transmon qubits coupled directly via a superconducting quantum interference device (SQUID) shunted by a capacitor, $C_\text{c}$.
This tuneable coupling scheme has recently been realised in circuit QED setups using transmons~\cite{kounalakis2018tuneable} and LC resonators~\cite{collodo2018observation}.
The coupling is controlled by tuning the Josephson energy of the SQUID, $E_\text{J}^\text{c}(\Phi_\text{b})=E_{\text{J},\Sigma}^\text{c}|\cos(\pi\Phi_\text{b}/\Phi_0)|$, with an out-of-plane flux bias, $\Phi_\text{b}$, where $\Phi_0=h/2e$ is the magnetic flux quantum and $E_{\text{J},\Sigma}^\text{c}$ is the sum of the two junction Josephson energies in the SQUID. 
The mechanical part of the circuit consists of a beam of length $l$ that is embedded in one of the arms of the SQUID loop and can oscillate out of plane.
By applying an in-plane external magnetic field $B$, the loop can pick up an additional flux $\beta_0 B l X$ due to the beam displacement $X$, resulting in a flux-tuneable and motion-dependent Josephson energy $E_\text{J}^\text{c}=E_{\text{J},\Sigma}^\text{c} |\cos(\pi\Phi_\text{b}/\Phi_0 + \alpha X)|$, where $\alpha=\pi\beta_0 B l/\Phi_0$ and $\beta_0$ is a geometric factor depending on the mode shape (for the first mechanical mode, considered here, $\beta_0\simeq1)$~\cite{etaki2008motion, shevchuk2017strong, rodrigues2019coupling}.

The Hamiltonian describing the circuit in Fig.~\ref{fig:circuit}a is
\begin{align}
H~=~& \frac{P^2}{2 m} +  \frac{m\omega_\text{m}^2X^2}{2}+\sum_{i=1}^2\left[\frac{Q_i^2}{2\widetilde{C}_i}- E_{\text{J}_i}\cos{\left(\frac{\phi_i}{\phi_0}\right)}\right] \nonumber\\
&+\frac{C_c}{C_1C_2}Q_1Q_2-E_\text{J}^\text{c}(X)\cos\left(\frac{\phi_1-\phi_2}{\phi_0}\right),
\label{eq:CircuitHam_main}
\end{align}
where $\{X,~P\}$ and $\{\phi_i,~Q_i\}$ are conjugate variable pairs describing the mechanical and the electrical degrees of freedom at circuit node $i$, and $\phi_0=\Phi_0/(2\pi)$ is the reduced flux quantum.
The first four terms describe the uncoupled system of the mechanical resonator and two transmon qubits, where $m,~\omega_\text{m}$ denote the mass and frequency of the beam and $E_{\text{J}_i}$, $\widetilde{C}_i$ represent the Josephson energy and loaded capacitance of each transmon, respectively (see Supplementary Sec.~S1).
The last two terms describe the qubit-qubit interaction via their charge and flux degrees of freedom, $Q_i$ and $\phi_i$.

The core of this proposal relies on the fact that the dynamical displacement of the beam results in a modulation of the superconducting current through the coupling SQUID and its Josephson energy, which mediates the qubit-qubit interaction.
Taking into account the finite asymmetry of the SQUID loop, which is present in any realistic scenario, this results in a motion-dependent Josephson energy,
\begin{equation}
E_\text{J}^\text{c}(X)=E_{\text{J},\Sigma}^\text{c} \left[c_\text{J}\cos(\pi\Phi_\text{b}/\Phi_0) - s_\text{J}\sin(\pi\Phi_\text{b}/\Phi_0) \alpha X\right],
\label{eq:EJx}
\end{equation}
where $c_\text{J}=\sqrt{1+a_\text{J}^2|\tan{(\pi\Phi_\text{b}/\Phi_0)}|}$ and $s_\text{J}=(1-a_\text{J}^2)/c_\text{J}$ are correction factors due to the SQUID asymmetry $a_\text{J}$.
Note that the above expression is valid for $\pi\Phi_\text{b}/\Phi_0 \gg \alpha X$ and relies on the assumption that $\alpha X\ll 1$ (for a full derivation see Supplementary Sec.~S1).
For the parameters considered here, which are compatible with values reported in recent experiments using micrometer-long Al beams and sub-Tesla magnetic fields~\cite{rodrigues2019coupling}, we have $\alpha X\sim10^{-5}-10^{-6}$, therefore this is a valid assumption.

\subsection{Electromechanical system dynamics}
The Hamiltonian of the system can be expressed in second quantisation form, as
\begin{align}
&\hat{H}~=~\hat{H}_0 + \hat{H}_\text{int},\\
&\hat{H}_0~=~\hbar\omega_\text{m}\hat{b}^\dagger\hat{b}+\sum_{i=1}^2\hbar\omega_i\hat{c}_i^\dagger\hat{c}_i-\frac{E_{\text{C}_i}}{2}\hat{c}_i^\dagger\hat{c}_i^\dagger\hat{c}_i\hat{c}_i,\\
&\hat{H}_\text{int}~=~\hbar g(\hat{c}_1^\dagger \hat{c}_2+\hat{c}_1\hat{c}_2^\dagger)(\hat{b}+\hat{b}^\dagger) \nonumber\\
&~~~~~~~~~~~~~~-\sum_{i=1}^2\hbar g_i\hat{c}_i^\dagger \hat{c}_i(\hat{b}+\hat{b}^\dagger)+\hat{H}_\text{12}^{'},
\label{eq:Ham_quantum}
\end{align}
where $\hat{b}^{(\dagger)}$ and $\hat{c}_i^{(\dagger)}$ are bosonic operators describing the annihilation (creation) of phonons and qubit excitations, respectively (see Supplementary Sec.~S2).
The effective electromechanical frequencies are $\omega_\text{m}$ and $\omega_i=\frac{1}{\hbar}\left(\sqrt{8\widetilde{E}_{\text{J}_i}E_{\text{C}_i}}-E_{\text{C}_i}\right)$, where $\widetilde{E}_{\text{J}_i}$ is the modified transmon Josephson energy due to the coupler.
The full quantum mechanical treatment of the circuit, including higher-order nonlinear interaction terms, is presented in the Supplementary Material.

The first term in Eq.~(\ref{eq:Ham_quantum}) describes a three-body interaction involving hopping of qubit excitations together with mechanical displacements of the beam.
The coupling strength is given by
\begin{equation}
g~=~\frac{\alpha\sqrt{Z_1Z_2}}{2\phi_0^2}s_\text{J}E_{\text{J},\Sigma}^\text{c}\sin(\pi\Phi_\text{b}/\Phi_0) X_\text{ZPF},
\end{equation}
where $Z_i=\frac{\hbar}{e^2}\sqrt{E_{\text{C}_i}/2\widetilde{E}_{\text{J}_i}}$ denote the transmon impedances, and $X_\text{ZPF}~=~\sqrt{{\hbar}/{(2m\omega_\text{m})}}$ is the zero-point motion of the mechanical resonator.
The next interaction term describes the radiation-pressure coupling of each qubit with the beam at a rate $g_{1(2)}~=~g \sqrt{Z_{1(2)}/Z_{2(1)}}$. 
The electromechanical coupling strengths are plotted in Fig.~\ref{fig:circuit}b as a function of the out-of-plane flux bias $\Phi_b$.
Close to a half-integer flux quantum, $E_\text{J}^\text{c}(X)$ is maximally susceptible to mechanical motion which maximises the electromechanical coupling.
Note that all couplings become zero exactly at a half-integer flux quantum due to the finite SQUID asymmetry $a_\text{J}$.
In our calculations an asymmetry of $0.01$ is included, reflecting a $2\%$ spread in junction fabrication targeting.

The last term in Eq.~(\ref{eq:Ham_quantum}) describes the qubit-qubit interaction
\begin{equation}
\hat{H}_\text{12}^{'}~=~\hbar (J_\text{C}-J_\text{L}) (\hat{c}_1^\dagger \hat{c}_2+\hat{c}_1\hat{c}_2^\dagger) + \hbar V \hat{c}_1^\dagger \hat{c}_1\hat{c}_2^\dagger \hat{c}_2,
\end{equation}
where $J_\text{C}$,~$J_\text{L}$ are exchange-type coupling strengths arising from the coupling capacitor and SQUID, respectively, and $V$ is the cross-Kerr coupling strength, which is minimised at $\Phi_b\simeq\Phi_0/2$ ~(see Supplementary Eq.~S(33)).
A significant advantage of this architecture for realising tripartite interactions, compared to relying exclusively on capacitive coupling~\cite{mehdi2015quantum}, is that any stray linear coupling between the qubits can be suppressed with the right choice of coupling capacitance $C_\text{c}$~\cite{kounalakis2018tuneable, collodo2018observation}.
This makes the three-body interaction dominant and ensures the ability to manipulate the state of each qubit individually by local driving, which is crucial for the state engineering protocols discussed below.

\subsection{Ground-state cooling}
Mechanical resonators realised using vibrating beams and drumheads lie in the radio-frequency regime ($\sim10$~MHz), where thermal fluctuations are dominant even at millikelvin temperatures, achieved with conventional cryogenic techniques.
An essential element of control is, therefore, the ability to cool these systems to their quantum ground state before manipulating them further.
Typically, electromechanical experiments employ a ``cold'' microwave cavity ($\sim10$~GHz), which is coupled to the mechanical element via an effective linearised interaction in the many-photon regime, and cooling is enabled by red-sideband driving~\cite{teufel2011sideband}.
However, in a system comprised of qubits, cooling the mechanical resonator via sideband driving can be a challenging task, requiring multiple tones and eventually limited by the critical number of photons in the Josephson junction~\cite{gely2019observation, lescanne2019observing}.

\begin{figure}[t]
  \begin{center}
  \includegraphics[width=1\linewidth]{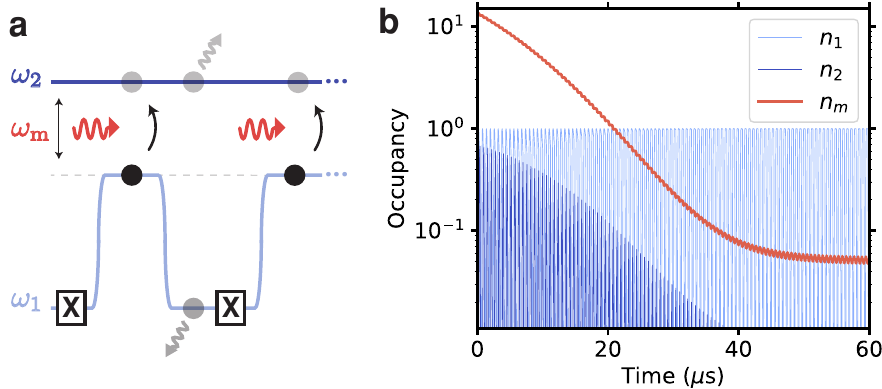}
  \end{center}
  \caption{{\bf Ground-state cooling.}
  {\bf a.} Schematic of the time-domain protocol to cool the mechanical resonator to its ground state using the three-body interaction. 
  In each cycle, qubit 1 is excited with a microwave pulse, then its frequency is tuned at $\omega_1=\omega_2-\omega_\text{m}$ for a variable time $\Delta t_\text{cool}$ followed by a reset on both qubits. 
  {\bf b.} Numerical results after $\sim100$ cycles demonstrating cooling to a 0.05 phonon occupancy for a 10~MHz resonator at $T=10$~mK ($n_\text{th}\simeq20$) using the system parameters presented in Table~\ref{tab:Parameters}.
}  \label{fig:Cooling}
\end{figure}

Here, we show that it is possible to overcome the challenges of sideband cooling with qubits by employing a time-domain protocol to cool the mechanical resonator to its quantum ground state, using the three-body interaction.
The scheme, depicted in Fig.~\ref{fig:Cooling}a, consists of a sequence of single-qubit operations which, combined with the tripartite interaction, enable the transfer of the thermal phonons to the environment in a stroboscopic fashion, as described below.
At first, we bring one qubit ($q_1$) to its excited state and then tune its frequency such that $\omega_1=\omega_2-\omega_\text{m}$.
Since $g\ll\omega_\text{m}$ the interaction $(\hat{c}_1\hat{b}\hat{c}_2^\dagger+\text{H.c.})$ is resonant at this condition, such that a phonon combined with the excitation in $q_1$ can be transferred to the other qubit ($q_2$) after variable time $\Delta t_\text{cool}$.
The cycle is then completed by reinitialising both qubits using active reset protocols~\cite{riste2012feedback, campagne2013persistent, magnard2018fast}.
A similar scheme has also been realised recently for detecting microwave photons in a superconducting resonator, using tripartite interactions with a transmon qubit and a dissipative mode~\cite{lescanne2019detecting}. 

In Fig.~\ref{fig:Cooling}b we plot the average number of phonons and qubit excitations as a function of time after~$\sim100$ cooling cycles, for a mechanical beam oscillating at $\omega_\text{m}/(2\pi)=10$~MHz and a tripartite coupling strength $g/(2\pi)\simeq0.3$~MHz.
At the end of the protocol the mechanical resonator is cooled down to the ground state with a phonon occupancy of 0.05, assuming an environment temperature of $T=10$~mK ($n_\text{th}\simeq20$).
The qubit ($\omega_1/(2\pi)=7$~GHz) is excited with a 200~ns Gaussian pulse, while the reset and cooling times are set to $\Delta t_\text{reset}=\Delta t_\text{cool}=200$~ns.
The cooling time would be best optimised by choosing a short time for large thermal occupation and then increase the time as the resonator cools, since phonons swap faster for higher occupations.
For simplicity, here, we considered a fixed time and found that for these parameters, a 200~ns time is sufficient.
Furthermore, we have assumed an in-plane magnetic field $B=10$~mT which is well-below the critical field for thin Al beams~\cite{meservey1971properties} and does not compromise the qubit coherence~\cite{schneider2019transmon}.
All system parameters used in the simulations are listed in Table~\ref{tab:Parameters}.

\begin{table}[h]
\begin{tabular}{l c }
\hline
\cline{1-2}
Parameter\ \ \ \ \ \ \ \ \  & Value\\ 
\hline
\hline
$\omega_\text{m}/(2\pi)$   & 10~MHz     \\
$\omega_{1,2}/(2\pi)$   & $\sim$7~GHz    \\
$g/(2\pi)$   & $ 0.3$~MHz     \\
$B$ & 10~mT     \\
$\Phi_\text{b}$    & $0.495~\Phi_0$       \\
$l~$ & 14.7~$\mu$m    \\
$X_\text{ZPF}$ & 33~fm \\
$\beta_0~$ & 1    \\
$C_{\text{c}}$    & 9.7~fF \\
$E_{\text{J},\Sigma}^\text{c}/E_{\text{J}_i}$   & $\sim10$ \\
$E_{\text{C}_i}/h$      & 320~MHz      \\
$n_\text{th}$~(T)   &  $\sim 20$~(10 mK)  \\
$T_1, T_2$  & 30~$\mu$s    \\
$Q_\text{m}$  & $10^6$      \\
\hline
\cline{1-2}
\end{tabular}
\caption{Parameter set used in the numerical simulations.}
 \label{tab:Parameters}
\end{table}

\subsection{Mechanical Fock states and qubit-phonon entanglement}
\begin{figure}[t]
  \begin{center}
\includegraphics[width=1\linewidth]{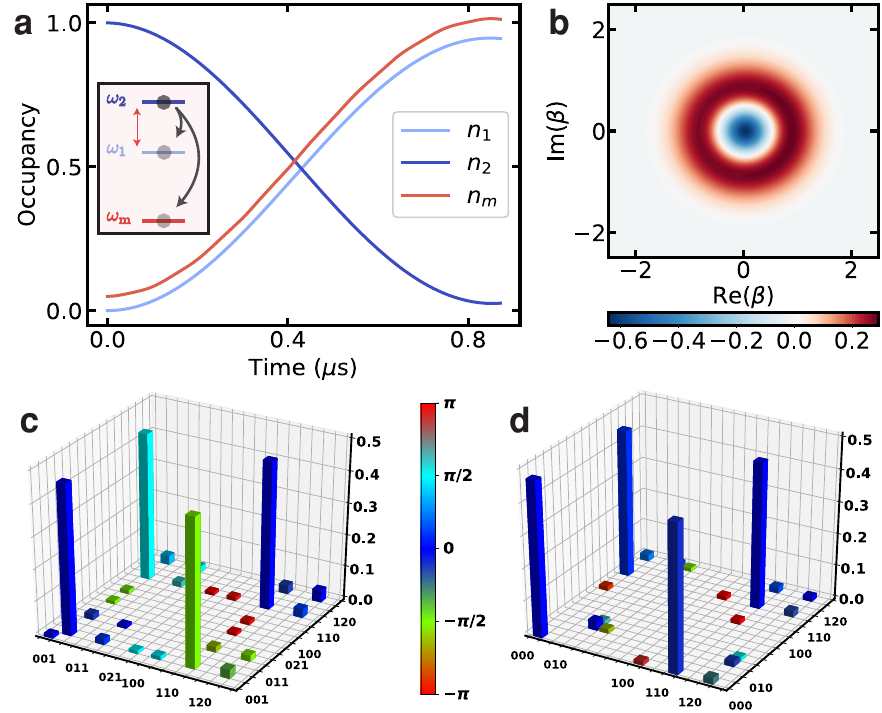}
  \end{center}
  \caption{{\bf Protocol for Fock-state preparation and maximally entangled states of phonons and qubit excitations}.
{\bf a.} Time evolution after exciting qubit 2 and tuning into the operating point $\omega_2=\omega_1+\omega_\text{m}$ (schematically presented in the inset) such that the interaction ($\hat{c}_1 b \hat{c}_2^\dagger+$H.c) is resonant.
Starting from a phonon occupancy of 0.05, a mechanical Fock state is generated with $97\%$ fidelity after time $t=\pi/(2g)$.
{\bf b.} Wigner function of the resulting mechanical state.
{\bf c.} Density matrix of a Greenberger--Horne--Zeilinger state that is generated in the middle of the protocol, at $t=\pi/(4g)$.
The notation $(0_10_\text{m}1_2)$ is used in labelling and only matrix elements with a magnitude larger than 0.005 are shown.
{\bf d.} Resulting density matrix after repeating the same protocol with qubit 2 initialised in the superposition state $\frac{1}{\sqrt{2}}(|0_2\rangle+|1_2\rangle)$, leading to a Bell state with $96\%$ fidelity.}  \label{fig:Fock}
\end{figure}

Following ground-state preparation, we present a protocol which employs the tripartite coupling to deterministically generate mechanical Fock states and maximally entangled states.
As schematically depicted in the inset of Fig.~\ref{fig:Fock}a, it consists of preparing $q_2$ in the excited state and tuning it to $\omega_2=\omega_1+\omega_\text{m}$, such that the interaction $(\hat{c}_1^\dagger\hat{b}^\dagger\hat{c}_2+\text{H.c.})$ is resonant.
Ideally, assuming unitary evolution $U(t)$ under the tripartite interaction, the state would evolve as
\begin{equation}
U(t)|0_10_\text{m}1_2\rangle= \cos(gt)|0_10_\text{m}1_2\rangle -i \sin(gt)|1_11_\text{m}0_2\rangle.
\end{equation}
The evolution of the average number of phonons and qubit excitations, starting from an attainable mechanical state of 0.05 thermal phonons and including system dissipation (see Methods) is plotted in Fig.~\ref{fig:Fock}a.
We consider the full interaction Hamiltonian presented in Eq.~(\ref{eq:Ham_quantum}), including next-to-leading order non-linear correction (see Supplementary Sec.~S2), for the simulation parameters shown in Table~\ref{tab:Parameters}.
In Fig.~\ref{fig:Fock}b we plot the Wigner function of the final state in the mechanical resonator, after time $t=\pi/(2g)$, revealing a single-phonon Fock state with $97\%$~($99\%$) fidelity, starting from an attainable (ideal) ground state.
Higher phonon states could also be prepared by resetting and repeating the protocol with modified transfer times $t_n=n_\text{m}^{-1/2}\pi/(2g)$.

The quantum state preparation scheme described above, can also be used to generate bipartite and tripartite maximally entangled states between the mechanical resonator and the qubits.
In particular, in the middle of the above protocol, at $t=\pi/(4g)$, the system is in a Greenberger--Horne--Zeilinger (GHZ) state 
\begin{equation}
|\psi\rangle_\text{f}=\frac{1}{\sqrt{2}}\left(|0_10_\text{m}1_2\rangle-i|1_11_\text{m}0_2\rangle\right),
\end{equation}
with $98~\%$ fidelity, with the corresponding density matrix shown in Fig.~\ref{fig:Fock}c.
Such states are particularly interesting for applications in quantum information~\cite{cleve1997substituting, bruss1998optimal} and fundamental tests of quantum theory~\cite{greenberger1989going}.
Using the same protocol for $q_2$ in a superposition state $\frac{1}{\sqrt{2}}(|0_2\rangle+|1_2\rangle)$, the Bell state
\begin{equation}
|\psi\rangle_\text{f}=\left(\frac{1}{\sqrt{2}}(|0_10_\text{m}\rangle+|1_11_\text{m}\rangle)\right)|0_2\rangle,
\end{equation}
is generated after time $\pi/(2g)$ with $96\%$ fidelity ($98\%$ for ideal ground state), as depicted in Fig.~\ref{fig:Fock}d.
The prepared state is a maximally entangled pair of a phonon and a qubit excitation, which could be utilised as a testbed for checking the validity of quantum mechanics at macroscopic scales without requiring tomography of the mechanical state~\cite{hofer2016proposal, vivoli2016proposal}.
Such states might also be suitable for integrating transmon qubits into other platforms such as spins, cold atoms, or even optical photons, via the mechanical resonator~\cite{rabl2010quantum, camerer2011realization, arcizet2011single, hill2012coherent}.
They could also provide possibilities for entangling the mechanical resonator with other physical systems via the transmon.

\subsection{Multi-phonon quantum superpositions}
\begin{figure*}[t]
  \begin{center}
  \includegraphics[width=0.7\linewidth]{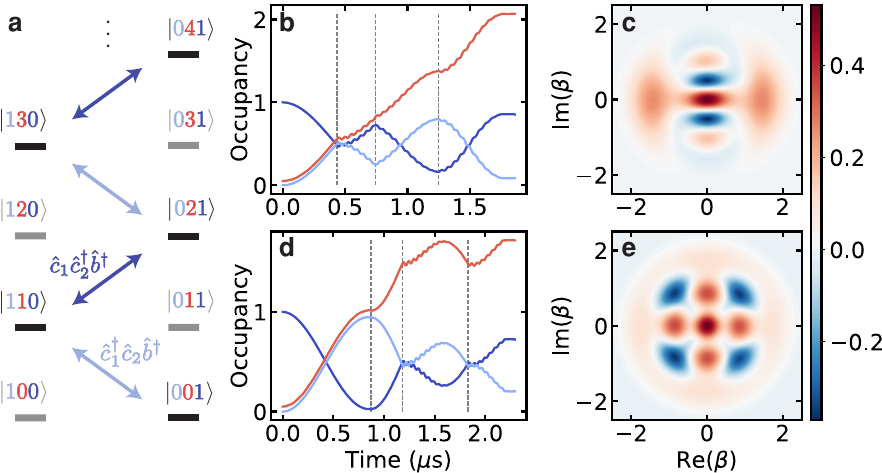}
  \end{center}
  \caption{{\bf Synthesizing multi-phonon quantum superposition states.}
  {\bf a.} Level diagram of the system indicating the resonant three-body interactions when the qubits are tuned such that $\omega_2^\pm~=~\omega_1\pm\omega_\text{m}$.
  {\bf b.},~{\bf d.} Evolution of phonon and qubit populations after exciting qubit 2 and alternating the qubit frequencies to $\omega_2^\pm$, where the dashed vertical lines denote the application of a square tuning flux pulse.
  An attainable mechanical state of 0.05 phonons is taken initially.
  {\bf c.},~{\bf e.} Wigner functions of the resulting states $\frac{1}{2}(|0\rangle+\sqrt{2}|2\rangle+|4\rangle)$ and $\frac{1}{\sqrt{2}}(|0\rangle+|4\rangle)$ with preparation fidelities $98\%$ and $97\%$ (following post-selection on $|0_11_2\rangle$), respectively.}
  \label{fig:Superpositions}
\end{figure*}

We now extend the protocol described above to create multi-phonon quantum superposition states in the mechanical resonator, simply by flux-pulsing the qubits.
In the protocols discussed previously, the qubit frequencies are tuned at $\omega_2=\omega_1+\omega_\text{m}$ such that the states $|0_1n_\text{m}1_2\rangle$ and $|1_1(n+1)_\text{m}0_2\rangle$ are coupled.
However, when tuned at $\omega_2=\omega_1-\omega_\text{m}$ the interaction term $(\hat{c}_1\hat{b}^\dagger\hat{c}_2^\dagger+\text{H.c.})$ becomes resonant, which couples $|0_1(n+1)_\text{m}1_2\rangle$ and $|1_1n_\text{m}0_2\rangle$.
Therefore, by interchanging the qubit frequencies with flux-tuning pulses during each cycle it could be possible to create higher phonon Fock states and multi-phonon quantum superposition states, as depicted schematically in Fig.~\ref{fig:Superpositions}a.

As a proof-of-concept, using the same simulation parameters as above (Table~\ref{tab:Parameters}), we demonstrate the creation of superposition states $|\psi_\text{m}\rangle^{'}=\frac{1}{2}(|0\rangle+\sqrt{2}|2\rangle+|4\rangle)$ and $|\psi_\text{m}\rangle^{''}=\frac{1}{\sqrt{2}}(|0\rangle+|4\rangle)$, after exciting qubit 2 and applying three flux pulses that interchange the qubit frequencies at variable times $t_1, t_2$ and $t_3$.
Figs.~\ref{fig:Superpositions}b,~d show the evolution of the qubit and resonator occupancy, starting from an attainable mechanical state of 0.05 phonons.
The dashed lines indicate the times that a flux-tuning pulse is applied.
The corresponding Wigner functions at the end of each protocol, following post-selection on $|0_11_2\rangle$, are shown in Figs.~\ref{fig:Superpositions}c,~e, with preparation fidelities $98~\%$ and $97~\%$, respectively.
After preparation, the states evolve naturally as $U(t)|\psi_\text{m}\rangle^{'}=(\alpha|0\rangle+\beta e^{-in\omega_\text{m}t}|n\rangle+\gamma e^{-im\omega_\text{m}t}|m\rangle)$ including dissipation, which is however not a limiting factor because of the long lifetimes of these mechanical resonators~\cite{rodrigues2019coupling}.
Readout of the prepared states including Wigner tomography could be performed using similar techniques to the ones developed in Ref.~\cite{hofheinz2009synthesizing}.

The scheme described above enables the generation of interesting classes of multi-phonon superposition states, such as the ones shown in Fig.~\ref{fig:Superpositions}, requiring only flux-tuning pulses and a projective measurement at the end.
As we show in Supplementary Sec.~S3, by including a projective measurement after each step of the protocol, it is possible to generate states with arbitrary phonon number probability distributions, although constrained in the relative phases of the superpositions.
Furthermore, we find that quantum superpositions with arbitrary complex coefficients can also be generated with this platform by additionally employing the qubit-qubit interaction in a controllable fashion to perform exchange-type and C-Phase gates between the two qubits (see Supplementary Sec.~S3).
This would enable the creation of truly arbitrary states, similar to those produced in resonantly coupled qubit-resonator systems~\cite{law1996arbitrary, hofheinz2009synthesizing}, with the trade-off of increased complexity in the protocol.
Alternatively, using the radiation-pressure coupling with one qubit in combination with a sequence of driving pulses, could enable the creation of mechanical Schr{\"o}dinger cat states as discussed in Ref.~\cite{mehdi2015quantum}.

\section{Discussion}
A reconciliation of quantum mechanics and general relativity remains elusive at a theoretical level, however, there exist several proposals for testing the quantum-classical boundary with mechanical resonators offering an ideal testbed.
More specifically, it has been theorised that a massive object in quantum superposition results in two coexisting space-time geometries, leading to issues with the unitary evolution, which eventually causes it to collapse~\cite{penrose1996gravity, kleckner2008creating}.
Importantly, this relies on the zero-point motion $X_\text{ZPF}$ being much larger than the approximate size of the nucleus ($\sim1$~fm), which is the case in our system ($X_\text{ZPF}=33$~fm).
The collapse timescale $t_\text{G}$ is inversely proportional to the mass of the object, resulting in $t_\text{G}\sim1-10$~s for the parameters considered here ($m\sim1$~pg), therefore the resonator coherence time should be larger than that.
Recent advances in strain engineering techniques can enable the enhancement of beam quality factors up to $Q_\text{m}\sim10^9$~\cite{ghadimi2018elastic}, leading to relaxation times of hundreds of seconds, which would be sufficient for observing gravitational effects.
Moreover, the ability to prepare a large variety of superposition states could offer an additional tool in testing such theories.
The proposed architecture provides a very versatile platform in this regard, enabling not only generic quantum state preparation, but also with high fidelity, which has so far been a very challenging task.

Our approach combines the best of both worlds of two very versatile systems, namely the exquisite level of quantum control of qubits in circuit quantum electrodynamics~\cite{blais2004cavity}, with the long lifetimes and flexibility of mechanical elements in coupling to electromagnetic radiation.
The high-fidelity generation of hybrid entangled states of phonons and qubit excitations, which have no classical analogue, may provide alternative routes for testing the limits of quantum theory at macroscopic scales~\cite{greenberger1989going, hofer2016proposal, vivoli2016proposal}.
Additionally, such states are of particular importance in enabling quantum technologies with hybrid quantum systems, from quantum simulation to quantum computing and communication~\cite{kurizki2015quantum}, and could also be used for coupling qubit excitations with other systems such as optical photons, cold atoms, or spin systems~\cite{rabl2010quantum, camerer2011realization, arcizet2011single, hill2012coherent}.

Furthermore, we have tested the robustness of our proposal against several imperfections that may occur in a realistic experimental scenario (see Supplementary Sec.~S4).
The most important limitation would be the presence of a considerable amount of flux noise, resulting in stray qubit-qubit coupling; for example, adding a fluctuation of $\delta\Phi_\text{b}=1-10~\mu\Phi_0$ results in $0.1-1$~MHz added qubit-qubit coupling $J$, respectively.
We find that for $J<1$~MHz the fidelity of the cooling and quantum state preparation protocols is not compromised (Supplementary Fig.~S4), therefore $\delta\Phi_\text{b}<10~\mu\Phi_0$ is required, which is compatible with observations in similar devices~\cite{kumar2016origin, hutchings2017tunable}.
We note that despite the steep slope of the tripartite coupling $g$ versus flux bias for $\Phi_\text{b}/\Phi_0>0.495$ (Fig.~\ref{fig:circuit}b), the tripartite coupling never changes by more than $1\%$ for $\delta\Phi_\text{b}<10~\mu\Phi_0$.
Another possible experimental limitation is the deviation from the target qubit frequencies due to imperfect flux tuning pulses.
We have studied the effect of this imperfection and find that targeting the qubit frequencies within 100~kHz is sufficient for high-fidelity quantum state preparation (see Supplementary Fig.~S5).
Additionally, we have studied the robustness of the protocol against qubit coherence and we find that high-fidelity quantum state preparation can be obtained for relaxation and dephasing times $T_{1,2}\gtrsim10~\mu$s, which are typical in the superconducting qubit community and compatible with 10~mT magnetic fields~\cite{schneider2019transmon}.

In conclusion, we have analysed a hybrid circuit architecture featuring strong and tuneable flux-mediated electromechanical interactions between a mechanical resonator and two superconducting transmon qubits.
Using state-of-the-art parameters, we find that the coupled system can operate in the single-photon strong-coupling regime, which has been a long-standing goal in the field of optomechanics.
We have proposed and numerically demonstrated several protocols for achieving ground-state cooling and preparing multi-phonon quantum superposition states as well as hybrid entanglement with high fidelities, which has been a tremendous challenge so far.
Moreover, the proposed schemes for quantum manipulation are applicable to a wider range of tripartite quantum systems where a lower frequency mode, that is not directly accessible, is within the tuning range of the two other controllable modes.
Our work significantly extends the quantum control toolbox of parametrically coupled radio-frequency mechanical resonators and provides a versatile on-chip interface with transmon-based processors, offering rich opportunities for technological applications as well as fundamental tests of quantum mechanics.

\section{Methods}
\subsection{Numerical modelling}
We model the dynamical evolution of the system, including environmental dissipation, with the Lindblad master equation
\begin{align}
\dot{\rho}~=~\frac{i}{\hbar}[\rho,\hat{H}]&+(n_\text{th}+1)\gamma_m\mathcal{L}[\hat{b}]\rho+n_\text{th}\gamma_m\mathcal{L}[\hat{b}^\dagger]\rho\nonumber\\
&+\sum_{i=1}^2\frac{1}{T_1}\mathcal{L}[\hat{c}_i]\rho+\frac{1}{T_2}\mathcal{L}[\hat{c}_i^\dagger\hat{c}_i]\rho,
\end{align}
which is numerically solved using QuTiP~\cite{johansson2012qutip}.
Here, $\mathcal{L}[\hat{o}]\rho\doteq(2\hat{o}\rho\hat{o}^\dagger-\hat{o}^\dagger\hat{o}\rho-\rho\hat{o}^\dagger\hat{o})/2$ are superoperators describing each dissipation process, and $n_\text{th}~=~1/[\exp(\hbar\omega_m/(k_\text{B}T))-1]$ is the thermal phonon number at temperature $T$.
More specifically, we consider qubit decay and dephasing times $T_1=T_2=30~\mu$s, which are consistent with measured values in a similar tuneable coupling transmon architecture~\cite{kounalakis2018tuneable}.
The coupling of the mechanical mode to the environment is determined by $\gamma_m=\omega_m/Q$, where the quality factor $Q=10^6$ is chosen in agreement with experimental observations in recently fabricated SQUID-embedded beams~\cite{rodrigues2019coupling}.
For completeness, we additionally include $\mathcal{O}(\phi^4X)$ terms in the interaction Hamiltonian~(see Supplementary Sec~S2), which nevertheless cause insubstantial corrections to the system dynamics.
We model the mechanical resonator using forty levels and each transmon as a three-level system (including an anharmonicity of $E_{\text{C}_i}/h\simeq320$~MHz).
The same parameters, shown in Table~\ref{tab:Parameters}, were considered in all the simulations.

\section{Acknowledgements}
We thank M.F. Gely, D. Bothner and I.C. Rodrigues for useful discussions.
This research was supported by the Dutch Foundation for Scientific Research (NWO) through the Casimir Research School.

\section{Author contributions}
M.K. conceptualized the work and developed the theory and numerical simulations with supervision from Y.M.B. and G.A.S.; M.K. wrote the manuscript with input from all coauthors.

\section{Data and code availability statement}
The simulation code and data sets generated and analysed during the current study
supporting the main and supplementary figures are publicly available in Zenodo with the identifier 10.5281/zenodo.3469853.

\cleardoublepage
\pagebreak
\newpage

\renewcommand{\theequation}{S\arabic{equation}}
\renewcommand{\thefigure}{S\arabic{figure}}
\renewcommand{\thetable}{S\arabic{table}}
\renewcommand{\bibnumfmt}[1]{[S#1]}
\onecolumngrid

\begin{center}
{\Large \textbf{Supplementary Information}}
\end{center}
\vspace{5pt}
\section{Lagrangian-Hamiltonian description of the circuit}
The Lagrangian describing the electromechanical system in Fig.1(a) is
\begin{align}
\mathcal{L}~=&~\frac{m\dot{X}^2}{2} -\frac{m\omega_\text{m}^2X^2}{2}+\sum_{i=1}^2\left[\frac{1}{2}C_i\dot{\phi}_i^2  +E_{\text{J}_i}\cos{\left(\frac{\phi_i}{\phi_0}\right)}\right]+\frac{1}{2}C_\text{c}(\dot{\phi}_1-\dot{\phi}_2)^2 + E_\text{J}^\text{c}\cos\left(\frac{\phi_1-\phi_2}{\phi_0}\right) ,
 \label{eq:Langragian}
\end{align}
where $X,~\phi_\text{i}$ are variables representing the beam displacement and the flux on circuit node $i$, respectively, and $\phi_0=\hbar/2e$ is the reduced flux quantum.
$C_i,~C_\text{c}$ and $E_{\text{J}_i},~E_\text{J}^\text{c}$ denote the capacitances and Josephson energies of each transmon and the coupler, respectively, and $m,~\omega_0$ are the mass and frequency of the beam. 
Following a Legendre transformation $H=\sum_i\phi_iQ_i-\mathcal{L}$ we obtain the system Hamiltonian
\begin{equation}
H~=~ \frac{P^2}{2 m} +  \frac{m\omega_\text{m}^2X^2}{2}+\sum_{i=1}^2\left[\frac{Q_i^2}{2\widetilde{C}_i}- E_{\text{J}_i}\cos{\left(\frac{\phi_i}{\phi_0}\right)}\right]+\frac{C_c}{C_1C_2}Q_1Q_2-E_\text{J}^\text{c}\cos\left(\frac{\phi_1-\phi_2}{\phi_0}\right),
\label{eq:CircuitHam}
\end{equation}
where $P~=~\frac{\partial \mathcal{L}}{\partial \dot{X}}$ is the mechanical conjugate momentum and $Q_i~=~\frac{\partial \mathcal{L}}{\partial \dot{\phi}_i}$ are the electrical conjugate momenta representing charges on each circuit node.
$\widetilde{C}_2=C_2+\frac{C_1C_\text{c}}{C_1+C_\text{c}}$ and $\widetilde{C}_1=C_1+\frac{C_2C_\text{c}}{C_2+C_\text{c}}$ denote the modified transmon capacitances due to the coupling capacitance $C_\text{c}$.

\subsection{Motion-dependent flux-tuneable Josephson energy}
If the two junctions of the SQUID are identical, the total Josephson energy is $E_\text{J}^\text{c}(\Phi_\text{b})=E_{\text{J},\Sigma}^\text{c}|\cos(\pi\Phi_\text{b}/\Phi_0)|$, where $E_{\text{J},\Sigma}^\text{c}$ is the sum of the Josephson energies of each junction.
In the presence of an in-plane magnetic field, $B$, the SQUID loop can pick up an additional flux, $\beta_0 B l X$, due the beam displacement, where $l$ is the length of the suspended arm and $\beta_0$ is a geometric constant associated with the mode shape of the beam (for the first mechanical mode $\beta_0\sim 1$)~\cite{etaki2008motion, rodrigues2019coupling}.
Therefore, the flux- and motion-dependent Josephson energy reads
\begin{equation}
E_\text{J}^\text{c}=E_{\text{J},\Sigma}^\text{c} |\cos(\pi\Phi_\text{b}/\Phi_0 + \alpha X)|,
\label{eq:EJc_noAsymm}
\end{equation}
where $\alpha~=~\pi\beta_0 B l/\Phi_0$ and $\Phi_0~=~h/2e$ is the magnetic flux quantum.
Using basic trigonometry and expanding to lowest order in $X$, assuming $\pi\Phi_\text{b}/\Phi_0 \gg \alpha X$, we find
\begin{equation}
E_\text{J}^\text{c}~\simeq~E_{\text{J},\Sigma}^\text{c} \left[\cos(\pi\Phi_\text{b}/\Phi_0)(1-\alpha^2 X^2) - \sin(\pi\Phi_\text{b}/\Phi_0) \alpha X\right],
\end{equation}
The above approximation holds for $\alpha X_\text{ZPF}\ll1$, which is valid for the beams considered in similar experiments and small magnetic fields ($B<1$~T)~\cite{rodrigues2019coupling}.

In the above, the case of a symmetric SQUID $E_\text{J,1}^\text{c}=E_\text{J,2}^\text{c}$ is considered, however in realistic devices a finite asymmetry needs to be taken into account.
The Josephson energy is therefore more accurately described by
\begin{equation}
E_\text{J}^\text{c}(\Phi_\text{b}, X)~=~E_{\text{J},\Sigma}^\text{c} [\cos^2(\pi\Phi_\text{b}/\Phi_0 + \alpha X)+a_\text{J}^2\sin^2(\pi\Phi_\text{b}/\Phi_0+\alpha X)]^{1/2},
\label{eq:EJc_exact}
\end{equation}
where $a_\text{J}=|({E_\text{J,1}^\text{c}-E_\text{J,2}^\text{c}})/({E_\text{J,1}^\text{c}+E_\text{J,2}^\text{c}})|$ is the SQUID asymmetry~\cite{koch2007charge}.
For $\pi\Phi_\text{b}/\Phi_0 \gg \alpha X$, we have
\begin{align}
\cos^2(\pi\Phi_\text{b}/\Phi_0 + \alpha X)\simeq\cos^2(\pi\Phi_\text{b}/\Phi_0 )-2\alpha X\cos(\pi\Phi_\text{b}/\Phi_0)\sin(\pi\Phi_\text{b}/\Phi_0),\nonumber\\
\sin^2(\pi\Phi_\text{b}/\Phi_0 + \alpha X)\simeq\sin^2(\pi\Phi_\text{b}/\Phi_0 )+2\alpha X\cos(\pi\Phi_\text{b}/\Phi_0)\sin(\pi\Phi_\text{b}/\Phi_0).
\end{align}
Substituting into Eq.~(\ref{eq:EJc_exact}), yields
\begin{equation}
E_\text{J}^\text{c}~=~E_{\text{J},\Sigma}^\text{c} [1+a_\text{J}^2\tan^2{(\pi\Phi_\text{b}/\Phi_0)}]^{1/2}\cos(\pi\Phi_\text{b}/\Phi_0) \left[1 - \alpha X\frac{(1-a_\text{J}^2)\tan(\pi\Phi_\text{b}/\Phi_0)}{1+a_\text{J}^2\tan^2{(\pi\Phi_\text{b}/\Phi_0)}}\right]^{1/2},
\end{equation}
which can further be simplified, assuming $\alpha X\frac{(1-a_\text{J}^2)\tan(\pi\Phi_\text{b}/\Phi_0)}{1+a_\text{J}^2\tan^2{(\pi\Phi_\text{b}/\Phi_0)}}\ll1$, into
\begin{equation}
E_\text{J}^\text{c}~\simeq~E_{\text{J},\Sigma}^\text{c} \left[c_\text{J}\cos(\pi\Phi_\text{b}/\Phi_0) - s_\text{J}\sin(\pi\Phi_\text{b}/\Phi_0) \alpha X\right]+\mathcal{O}[X^2],
\label{eq:EJc_expanded}
\end{equation}
where $c_\text{J}=\sqrt{1+a_\text{J}^2\tan^2{(\pi\Phi_\text{b}/\Phi_0)}}$ and $s_\text{J}=(1-a_\text{J}^2)/c_\text{J}$.
The higher order terms
\begin{equation} 
\mathcal{O}[X^2] = E_{\text{J},\Sigma}^\text{c}\frac{s_\text{J}\sin^2(\pi\Phi_\text{b}/\Phi_0)}{2c_\text{J}\cos(\pi\Phi_\text{b}/\Phi_0)}\alpha^2X^2,
\end{equation}
are considered in the analysis below, however, as we will see later, they do not affect the system dynamics for the parameters used in this work.

\subsection{Flux-mediated interactions}
Expanding the last term in Eq.~(\ref{eq:CircuitHam}) up to $\mathcal{O}[\phi^4X^2]$, in combination with Eq.~(\ref{eq:EJc_expanded}), yields the following flux-mediated interaction terms
\begin{align}
H_\text{int}^\text{(flux)}~=&~H_\text{3-body} + H_\text{RP} + H^{\{\phi^2\}} + H^{\{\phi^4\}} + H^{\{\phi^4X\}}+ H^{\{\phi^2X^2\}}  +  H^{\{\phi^4X^2\}}.
\label{eq:Ham_int}
\end{align}

The first term describes a three-body interaction, between the two qubits and the beam
\begin{equation}
H_\text{3-body}~=~ \alpha~E_{\text{J},\Sigma}^\text{c} s_\text{J}\sin(\pi\Phi_\text{b}/\Phi_0)~\frac{\phi_1 \phi_2}{\phi_0^2} X,
\label{eq:Ham_3-body}
\end{equation}
while the second term
\begin{equation}
H_\text{RP}~=~-\alpha~E_{\text{J},\Sigma}^\text{c}s_\text{J} \sin(\pi\Phi_\text{b}/\Phi_0)~\frac{(\phi_1^2 + \phi_2^2) }{2\phi_0^2} X,
\label{eq:Ham_RP}
\end{equation}
describes a radiation-pressure type coupling of the mechanical mode with each qubit.

The next two terms describe flux-mediated qubit-qubit interactions, where
\begin{equation}
H^{\{\phi^2\}}~=~- E_{\text{J},\Sigma}^\text{c}c_\text{J} \cos(\pi\Phi_\text{b}/\Phi_0)~\frac{\phi_1 \phi_2}{\phi_0^2},
\label{eq:Ham_qr_flux}
\end{equation}
is a linear dipole coupling term, and
\begin{equation}
H^{\{\phi^4\}}= E_{\text{J},\Sigma}^\text{c} c_\text{J} \cos(\pi\Phi_\text{b}/\Phi_0) \left(-\frac{\phi_1^2 \phi_2^2}{4\phi_0^4}+\frac{\phi_1^3\phi_2 + \phi_1\phi_2^3}{6\phi_0^4}\right),
\label{eq:Ham_phi4}
\end{equation}
is a nonlinear interaction including cross-Kerr as well as pair- and correlated-hopping terms.

The last three terms
\begin{equation}
H^{\{\phi^4X\}}~=~\alpha~E_{\text{J},\Sigma}^\text{c} s_\text{J}\sin(\pi\Phi_\text{b}/\Phi_0)~\frac{(\phi_1 - \phi_2)^4 }{24\phi_0^4} X,
\end{equation}
\begin{equation}
H^{\{\phi^2X^2\}}~=~-\alpha^2~E_{\text{J},\Sigma}^\text{c}\frac{s_\text{J}\sin^2(\pi\Phi_\text{b}/\Phi_0)}{2c_\text{J}\cos(\pi\Phi_\text{b}/\Phi_0)}~\frac{\left(\phi_1-\phi_2\right)^2 }{2\phi_0^2}X^2,
\label{eq:tripartite_crossKerr}
\end{equation}
and
\begin{equation}
H^{\{\phi^4X^2\}}~=~\alpha^2~E_{\text{J},\Sigma}^\text{c}\frac{s_\text{J}\sin^2(\pi\Phi_\text{b}/\Phi_0)}{2c_\text{J}\cos(\pi\Phi_\text{b}/\Phi_0)}~\frac{(\phi_1 - \phi_2)^4}{24\phi_0^4} X^2,
\end{equation}
describe nonlinear tripartite interactions, which are much weaker and do not contribute significantly to the dynamics compared to the leading order electromechanical terms of Eq.~(\ref{eq:Ham_3-body}) and (\ref{eq:Ham_RP}).

In addition, the last term in Eq.~(\ref{eq:CircuitHam}) leads to corrections in the bare subsystem Hamiltonians.
More specifically, the inductive energy term
\begin{equation}
E_{\text{J},\Sigma}^\text{c} c_\text{J}\cos(\pi\Phi_\text{b}/\Phi_0) \left(\frac{\phi_1^2 + \phi_2^2}{2\phi_0^2}-\frac{\phi_1^4 + \phi_2^4}{24\phi_0^4}\right),
\label{eq:H0corr}
\end{equation}
results in an effective qubit Josephson energy 
\begin{equation}
\widetilde{E}_{\text{J}_i}~=~E_{\text{J}_i}+E_{\text{J},\Sigma}^\text{c}c_\text{J}\cos(\pi\Phi_\text{b}/\Phi_0),
\end{equation}
while the potential energy term
\begin{equation}
E_{\text{J},\Sigma}^\text{c}\alpha X s_\text{J}\sin(\pi\Phi_\text{b}/\Phi_0)
\label{eq:H0corrX}
\end{equation}
leads to a displaced rest position
\begin{equation}
X_0~=~\frac{\alpha E_{\text{J},\Sigma}^\text{c}s_\text{J} \sin(\pi\Phi_\text{b}/\Phi_0)}{2m\omega_\text{m}}.
\end{equation}
The latter does not affect the dynamics and can be absorbed in a redefinition of $X\rightarrow (X-X_0)$.

\section{Circuit quantisation}
We now switch to a quantum mechanical description of the circuit, promoting all canonical variables to quantum operators
\begin{align}
&\hat{X}~=~X_\text{ZPF}~(\hat{b}+\hat{b}^\dagger),~\hat{P}~=~P_\text{ZPF}~i(\hat{b}^\dagger-\hat{b}),\nonumber\\
&\hat{\phi}_i~=~\sqrt{\frac{\hbar Z_i}{2}}~(\hat{c}_i+\hat{c}_i^\dagger),~\hat{Q}_i~=~\sqrt{\frac{\hbar}{2Z_i}}~i(\hat{c}_i^\dagger-\hat{c}_i),
\end{align}
where $\hat{c}_i^{(\dagger)}$ and $\hat{b}^{(\dagger)}$ are ladder operators describing the annihilation (creation) of photons and phonons, respectively, and satisfy bosonic commutation relations $[\hat{c}_i,~\hat{c}_j^\dagger]~=~\delta_{ij}$ and $[\hat{b},~\hat{b}^\dagger]~=~1$. 
The zero-point fluctuations in the mechanical displacement and momentum are given by $X_\text{ZPF}~=~\sqrt{{\hbar}/{(2m\omega_\text{m})}}$ and $P_\text{ZPF}~=~\sqrt{{\hbar m\omega_\text{m}}/{2}}$, respectively.
$Z_i=\frac{\hbar}{e^2}\sqrt{E_{\text{C}_i}/2\widetilde{E}_{\text{J}_i}}$ denotes the impedance of each transmon, where $E_{\text{C}_i}~=~\frac{e^2}{2\widetilde{C}_i}$ is its charging energy.
Since the qubits are in the transmon regime~\cite{koch2007charge}, $\widetilde{E}_{\text{J}_i}\gg E_{\text{C}_i}$, the uncoupled electromechanical system is well-described by
\begin{align}
\hat{H}_0~=&~\hbar\omega_\text{m}\hat{b}^\dagger\hat{b}+\sum_{i=1}^2\hbar\omega_i\hat{c}_i^\dagger\hat{c}_i-\frac{E_{\text{C}_i}}{2}\hat{c}_i^\dagger\hat{c}_i^\dagger\hat{c}_i\hat{c}_i,
\end{align}
with qubit frequencies $\omega_i~=~\frac{1}{\hbar}\left(\sqrt{8\widetilde{E}_{\text{J}_i}E_{\text{C}_i}}-E_{\text{C}_i}\right)$.
\\

\subsection{Tripartite coupling}
Following Eq.~(\ref{eq:Ham_3-body}) we can now write the Hamiltonian operator describing the three-body interaction
\begin{align}
\hat{H}_\text{3-body}~&=~\hbar g~(\hat{c}_1+\hat{c}_1^\dagger)(\hat{c}_2+\hat{c}_2^\dagger)(\hat{b}+\hat{b}^\dagger),\nonumber\\
&=~\hbar g~(\hat{c}_1^\dagger \hat{c}_2+\hat{c}_1\hat{c}_2^\dagger)(\hat{b}+\hat{b}^\dagger),
\end{align}
where, in the second step, we have made a rotating-wave approximation (RWA) to neglect fast rotating terms $\hat{c}_\text{1} \hat{c}_\text{2}\hat{b}^{(\dagger)} $ and $\hat{c}_\text{1}^{\dagger} \hat{c}_\text{2}^{\dagger}\hat{b}^{(\dagger)} $, which do not contribute to the dynamics since $\omega_1,\omega_2\gg \omega_\text{m}, g$.
The tripartite coupling strength is given by
\begin{align}
g~=~\frac{\alpha\sqrt{Z_1Z_2}}{2\phi_0^2}E_{\text{J},\Sigma}^\text{c}\sin(\pi\Phi_\text{b}/\Phi_0) X_\text{ZPF}.
\end{align}
\\

\subsection{Radiation-pressure couplings}
The radiation-pressure interaction between each qubit and the beam, in Eq.~(\ref{eq:Ham_RP}), can be expressed in second quantisation form as
\begin{align}
\hat{H}_\text{RP}~=~\hbar g_1~\hat{c}_1^\dagger \hat{c}_1(\hat{b}+\hat{b}^\dagger)~+~\hbar g_2~\hat{c}_2^\dagger \hat{c}_2(\hat{b}+\hat{b}^\dagger),
\end{align}
following a RWA.
The radiation-pressure coupling strengths are given by
\begin{align}
g_i~=~\frac{\alpha Z_i}{2\phi_0^2} E_{\text{J},\Sigma}^\text{c}\sin(\pi\Phi_\text{b}/\Phi_0) X_\text{ZPF}.
\end{align}
\\

\subsection{Qubit-qubit couplings}
Following circuit quantisation and a RWA, the linear qubit-qubit interaction Hamiltonian in Eq.~(\ref{eq:Ham_qr_flux}) becomes
\begin{equation}
H^{\{\phi^2\}}~=~-\hbar J_\text{L}~(\hat{c}_1^\dagger\hat{c}_2+\hat{c}_1\hat{c}_2^\dagger)
\end{equation}
where
\begin{align}
J_\text{L}~=~\frac{\sqrt{Z_1Z_2}}{2\phi_0^2}E_{\text{J},\Sigma}^\text{c} \cos(\pi\Phi_\text{b}/\Phi_0),
\end{align}
The qubits also couple via their charge degrees of freedom ($Q_1Q_2$ term in Eq.~(\ref{eq:CircuitHam})), which results in the same type of linear interaction with coupling strength
\begin{align}
J_\text{C}~=~\frac{C_\text{c}}{2C_1C_2} \left(Z_1Z_2\right)^{-1/2}.
\end{align}
Combining these together leads to an overall exchange-type interaction
\begin{align}
H_{12}~=~\hbar (J_\text{C}-J_\text{L})~(\hat{c}_1^\dagger\hat{c}_2+\hat{c}_1\hat{c}_2^\dagger),
\label{eq:Jqr}
\end{align}
which can be suppressed at the desired operating point ($\Phi_\text{b}$) with the right choice of coupling capacitor $C_\text{c}$.

The nonlinear interaction between the two transmons in Eq.~(\ref{eq:Ham_phi4}) is given by
\begin{align}
\hat{H}^{\{\phi^4\}}~=&~\frac{\hbar V}{4}(\hat{c}_1^\dagger+\hat{c}_1)^2(\hat{c}_2^\dagger+\hat{c}_2)^2\nonumber\\
&+\frac{\hbar J_{n_1}}{3}(\hat{c}_1^\dagger+\hat{c}_1)^3(\hat{c}_2^\dagger+\hat{c}_2) \nonumber\\
&+ \frac{\hbar J_{n_2}}{3}(\hat{c}_1^\dagger+\hat{c}_1)(\hat{c}_2^\dagger+\hat{c}_2)^3.
\label{eq:Hphi4}
\end{align}
The first term, following a RWA, results in a cross-Kerr interaction, $V \hat{c}_1^\dagger\hat{c}_1\hat{c}^\dagger_2\hat{c}_2$, with coupling strength
\begin{equation}
V~=~-\frac{\hbar Z_1Z_2}{4\phi_0^4}E_{\text{J},\Sigma}^\text{c} \cos(\pi\Phi_\text{b}/\Phi_0).
\end{equation}
Additionally the same term yields a pair-hopping interaction $\frac{V}{4}(\hat{c}_1^\dagger\hat{c}_1^\dagger\hat{c}_2\hat{c}_2+\text{H.c.})$, which however is not contributing to the dynamics for single-excitation levels.
Finally, the other two terms in Eq.~(\ref{eq:Hphi4}) result in correlated hopping interactions $[\hat{c}_1^\dagger (J_{n_1}\hat{c}_1^\dagger\hat{c}_1+J_{n_2}\hat{c}_2^\dagger\hat{c}_2)\hat{c}_2 + \text{H.c.}]$, with
\begin{equation}
J_{n_{1(2)}}~=~-\frac{\hbar \sqrt{Z_{1(2)}^3Z_\text{{2(1)}}}}{24\phi_0^4}E_{\text{J},\Sigma}^\text{c} \cos(\pi\Phi_\text{b}/\Phi_0),
\end{equation}
which contribute to the linear coupling in Eq.~(\ref{eq:Jqr}) as $J\rightarrow J+J_{n_1}+J_{n_2}$.
\\
\subsection{Higher-order tripartite interactions} 
The next-to-leading order electromechanical interactions are given by
\begin{align}
\hat{H}^{\{\phi^4X\}}~&=~\hbar g_{\phi_1^2\phi_2^2 x}(\hat{c}_1^\dagger+\hat{c}_1)^2(\hat{c}_2^\dagger+\hat{c}_2)^2(\hat{b}+\hat{b}^\dagger)\nonumber\\
&+\hbar g_{\phi_1^3\phi_2 x}(\hat{c}_1^\dagger+\hat{c}_1)^3(\hat{c}_2^\dagger+\hat{c}_2)(\hat{b}+\hat{b}^\dagger) \nonumber\\
&+\hbar g_{\phi_1\phi_2^3 x}~(\hat{c}_1^\dagger+\hat{c}_1)(\hat{c}_2^\dagger+\hat{c}_2)^3(\hat{b}+\hat{b}^\dagger),
\end{align}
with nonlinear coupling strengths
\begin{align}
g_{\phi_1^2\phi_2^2 x}~=~\frac{\hbar\alpha Z_1Z_2}{16\phi_0^4}E_{\text{J},\Sigma}^\text{c}\sin(\pi \Phi_\text{b}/\Phi_0) X_\text{ZPF},
\end{align}
\begin{align}
g_{\phi_1^3\phi_2 x}~&=~\frac{\hbar\alpha Z_1^{3/2}Z_2^{1/2}}{24\phi_0^4}E_{\text{J},\Sigma}^\text{c}\sin(\pi \Phi_\text{b}/\Phi_0) X_\text{ZPF},
\end{align}
and
\begin{align}
g_{\phi_1\phi_2^3 x}~=~\frac{\hbar\alpha Z_1^{1/2}Z_2^{3/2}}{24\phi_0^4}E_{\text{J},\Sigma}^\text{c}\sin(\pi \Phi_\text{b}/\Phi_0) X_\text{ZPF}.
\end{align}
The last two terms also lead to a small correction of the tripartite coupling strength $g\rightarrow g+3g_{\phi_1^3\phi_2 x}+3g_{\phi_1\phi_2^3 x}$.
These terms, although weaker, are not negligible (since they are of the same order or larger than dissipation rates) and they are include in simulating the system dynamics.

The $\mathcal{O}[\phi^2 X^2]$ terms in Eq.~(\ref{eq:tripartite_crossKerr}) can be written as, 
\begin{equation}
\hat{H}^{\{\phi^2X^2\}}~=~ \Sigma_{i,j} \hbar g_{\phi_i\phi_j x^2}\hat{c}_i^\dagger\hat{c}_j(\hat{b}+\hat{b}^\dagger)^2
\end{equation}
where 
\begin{equation}
g_{\phi_i\phi_j x^2}=\frac{\alpha^2\sqrt{Z_iZ_j}}{2\phi_0^2}E_{\text{J},\Sigma}^\text{c}\frac{s_\text{J}\sin^2(\pi\Phi_\text{b}/\Phi_0)}{2c_\text{J}\cos(\pi\Phi_\text{b}/\Phi_0)}X^2_\text{ZPF}.
\end{equation}
For the parameters used in this work ($\alpha X_\text{ZPF}\sim10^{-6}$) the coupling is negligible and does not affect the system dynamics (the same holds for the even weaker $\mathcal{O}[\phi^4 X^2]$ terms).
Interestingly, however, for $i=j$ we find a dispersive (cross-Kerr) interaction $\hat{c}_i^\dagger\hat{c}_i\hat{b}^\dagger\hat{b}$ of each qubit with the resonator, which is active even when the qubits are far detuned and could potentially be employed for phonon-sensitive measurements of the mechanical state~\cite{viennot2018phonon}.

\section{Protocols for arbitrary quantum state generation}
\subsection{States with arbitrary complex coefficients}
In the main text we demonstrated a protocol for creating multi-phonon quantum superposition states by alternating the qubit frequencies such that different parts of the tripartite interaction become resonant.
By controlling the interaction times and post-selecting on the qubit state at the end of the protocol, we showed the possibility of creating interesting classes of superposition states with high fidelity.
However, this protocol alone is not sufficient for generating mechanical superposition states $|\psi\rangle~=~\sum_{n=0}^N c_n|n_\text{m}\rangle$ with arbitrary complex coefficients $c_n$, since there is no phase degree of freedom to control in the protocol.
This is due to the fact that the coupling constant $g$ is not an adjustable complex number and in practical implementations the interaction time is the only parameter that can be varied.

However, there is another tuning knob that can be employed by taking advantage of the qubit-qubit exchange-type interaction $J~(\hat{c}_1^\dagger\hat{c}_2+\hat{c}_1\hat{c}_2^\dagger)$.
Detuning the qubits by $\Delta=|\omega_1-\omega_2|\gg\omega_\text{m}$, while tuning the coupling SQUID such that there is a finite exchange-type coupling strength $J\geq\Delta$, results in a resonant qubit-qubit interaction that couples $|0_11_2\rangle$ and $|1_10_2\rangle$.
For example, starting from $|0_11_2\rangle$ one would end up with state $[\cos(J t)|0_11_2\rangle-i\sin(Jt)|1_10_2\rangle]$ after time $t$ (at $t=\pi/(2J)$ this realises a SWAP gate).
Furthermore, by detuning the qubit frequencies such that $\Delta>J$ it is also possible to introduce a relative phase between the qubit states $|0_11_2\rangle$ and $|1_10_2\rangle$ (C-Phase gate).
Combining this gate with the resonant qubit-qubit interaction, acting on $|0_11_2\rangle$ would result in
\begin{equation}
U_\text{J}(t,\theta)|0_11_2\rangle ~=~\cos(J t)|0_11_2\rangle-ie^{i\theta}\sin(Jt)|1_10_2\rangle.
\label{eq:Uj}
\end{equation}

We note that our system, is analogous to the one studied theoretically by Law and Eberly~\cite{law1996arbitrary} and experimentally by Hofheinz et al.~\cite{hofheinz2009synthesizing}.
The system studied in these references considers a resonator that is controllably coupled to a qubit (with local qubit driving) via a resonant exchange-type interaction (Jaynes-Cummings).
It is shown that an arbitrary resonator state can be generated by interleaving the Jaynes-Cummings evolution with qubit driving.
The two-qubit states $|0_11_2\rangle$ and $|1_10_2\rangle$ in our system can be mapped to the single qubit basis $|0\rangle$ and $|1\rangle$ considered in Law and Eberly. Furthermore, the tripartite interaction can be mapped to the qubit-resonator Jaynes-Cummings interaction, where in our system we additionally have the possibility of realising the equivalent of a counter-rotating Jaynes-Cummings interaction by exchanging the qubit frequencies.
The equivalent of qubit driving can then be performed by controlling the evolution under the exchange qubit-qubit interaction $U_\text{J}$ described in Eq.~(\ref{eq:Uj}).

\begin{figure}[t]
    \centering
  \includegraphics[width=1\linewidth]{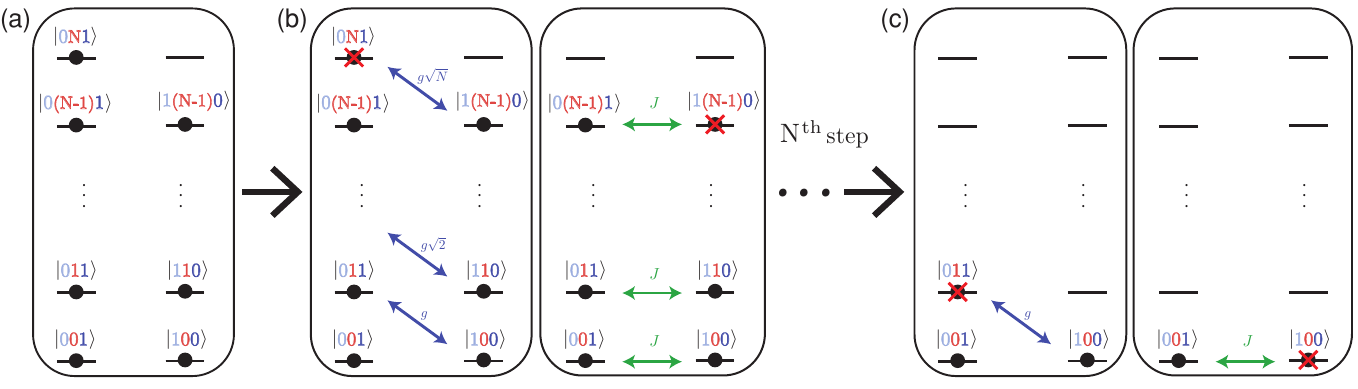}
  \caption{ {\bf Protocol for arbitrary state generation.}
  (a) The protocol is calculated in reverse by emptying an arbitrary entangled state $|\psi\rangle=\sum_{n=0}^{N-1} (a_n|0_1n_\text{m}1_2\rangle+b_n|1_1n_\text{m}0_2\rangle)+a_N|0_1N_\text{m}1_2\rangle$ (from this state it is always possible to obtain an arbitrary mechanical state by post-selecting on $|0_11_2\rangle$).
  (b) The first step of the protocol is shown, which consists of two substeps:
  At first we turn on the (inverse of the) C-Phase gate and the tripartite interaction $\hat{U}(t)=e^{i(\hat{c}_1^\dagger \hat{b} \hat{c}_2+\text{H.c})t}$ such that all population from $|0_1N_\text{m}1_2\rangle$ is emptied to $|1_1(N-1)_\text{m}0_2\rangle$.
  Then, by applying another C-Phase gate in combination with resonant qubit-qubit interaction for a variable time it is possible to empty $|1_1(N-1)_\text{m}0_2\rangle$ to $|0_1(N-1)_\text{m}1_2\rangle$.
  (c) Following this procedure $N$ times it is possible to empty the original state in (a) and end up in $|0_10_\text{m}1_2\rangle$.
 }
  \label{fig:ArbitraryStatesEmpty}
\end{figure}

We will consider the case where an arbitrary state $|\psi\rangle_\text{m}=\sum_{n=0}^{N} a_n|n_\text{m}\rangle$ is generated following post-selection from the arbitrary entangled state $|\psi\rangle=\sum_{n=0}^{N-1} (a_n|0_1n_\text{m}1_2\rangle+b_n|1_1n_\text{m}0_2\rangle)+a_N|0_1N_\text{m}1_2\rangle$.
That is, if we can create the above state $|\psi\rangle$ with arbitrary complex coefficients, it is then straightforward to collapse it to $|\psi\rangle_\text{m}$ following post-selection on the $|0_11_2\rangle$ qubit state.
Inspired by Refs.~\cite{law1996arbitrary, hofheinz2009synthesizing}, let us now consider the problem of generating the arbitrary state $|\psi\rangle$ from the inverse point of view, i.e. by proving that it is always possible to empty it.
Suppose we start from $|\psi\rangle=\sum_{n=0}^{N-1} (a_n|0_1n_\text{m}1_2\rangle+b_n|1_1n_\text{m}0_2\rangle)+a_N|0_1N_\text{m}1_2\rangle$ (Fig.~\ref{fig:ArbitraryStatesEmpty}a).
Then by applying the (inverse of the) C-Phase gate and the tripartite interaction $\hat{U}^\dagger(t)=e^{i(\hat{c}_1^\dagger \hat{b} \hat{c}_2+\text{H.c})t}$ for phase $\theta_\text{A}$ and time $t_\text{A}$, the probability amplitude of state $|0_1N_\text{m}1_2\rangle$ becomes
\begin{equation}
a_N\rightarrow a_N\cos{(g\sqrt{N}t_\text{A})}+ie^{\theta_\text{A}}b_{N-1}\sin{(g\sqrt{N}t_\text{A})}.
\end{equation}
By appropriately choosing $\theta_\text{A}$ and $t_\text{A}$, it is possible to make the above probability amplitude zero, such that all phonons occupying the state $|0_1N_\text{m}1_2\rangle$ are transferred to $|1_1(N-1)_\text{m}0_2\rangle$ (see Fig.~\ref{fig:ArbitraryStatesEmpty}b).

The step above also results in incomplete transfer of phonons between states $|0_1n_\text{m}1_2\rangle$ and $|1_1(n-1)_\text{m}0_2\rangle$ (for $n<N$), after which we are left with the state $|\psi\rangle^\prime=\sum_{n=0}^{N-1} (a^\prime_n|0_1n_\text{m}1_2\rangle+b^\prime_n|1_1n_\text{m}0_2\rangle)$.
Now by combining the (inverse of the) C-Phase gate and resonant qubit-qubit interaction $U_\text{J}^\dagger(t_\text{J},\theta_\text{J})|\psi\rangle^\prime$ the probability amplitude of state $|1_1(N-1)_\text{m}0_2\rangle$ becomes
\begin{equation}
b_{N-1}\rightarrow b_{N-1}\cos{(Jt_\text{J})}+ie^{\theta_\text{J}}a_{N-1}\sin{(Jt_\text{J})},
\end{equation}
which can be made zero by appropriately choosing $\theta_\text{J}$ and $t_\text{J}$ such that all phonons occupying $|1_1(N-1)_\text{m}0_2\rangle$ are transferred to $|0_1(N-1)_\text{m}1_2\rangle$ (see Fig.~\ref{fig:ArbitraryStatesEmpty}b).
By applying the above two steps $N$ times it is possible to completely empty the initial state, leading the system to $|0_10_\text{m}1_2\rangle$ (Fig.~\ref{fig:ArbitraryStatesEmpty}c).
Therefore, reversing the problem, it is possible to generate any arbitrary mechanical state $|\psi\rangle_\text{m}=\sum_{n=0}^{N} a_n|n_\text{m}\rangle$ by creating the arbitrary entangled state
\begin{equation}
\prod_{j=N}^1\hat{U}_{{\theta_\text{A}}_j}\hat{U}({t_A}_j)\hat{U}_\text{J}({t_\text{J}}_j, {\theta_\text{J}}_j)|0_10_\text{m}1_2\rangle~=~\sum_{n=0}^{N-1} (a_n|0_1n_\text{m}1_2\rangle+b_n|1_1n_\text{m}0_2\rangle)+a_N|0_1N_\text{m}1_2\rangle,
\end{equation}
and post-selecting on $|0_11_2\rangle$.

\subsection{States with arbitrary phonon number probability distributions}

The protocol presented in the previous section can enable the creation of any mechanical quantum state with arbitrary coefficients.
Although this protocol is experimentally feasible, it can become complex as it requires a lot of additional tuning to realise the C-Phase and exchange-type gates between the qubits.
Here we describe an alternative protocol that only employs the tripartite interactions, therefore requiring only alternating between the qubit frequencies, and post-selective measurements at each step.
Reducing the complexity of tuning pulses comes at the cost of not being able to create states with arbitrary complex coefficients, although it is possible to generate states with arbitrary phonon number probability distributions as we see below.

\begin{figure}[t]
    \centering
  \includegraphics[width=0.8\linewidth]{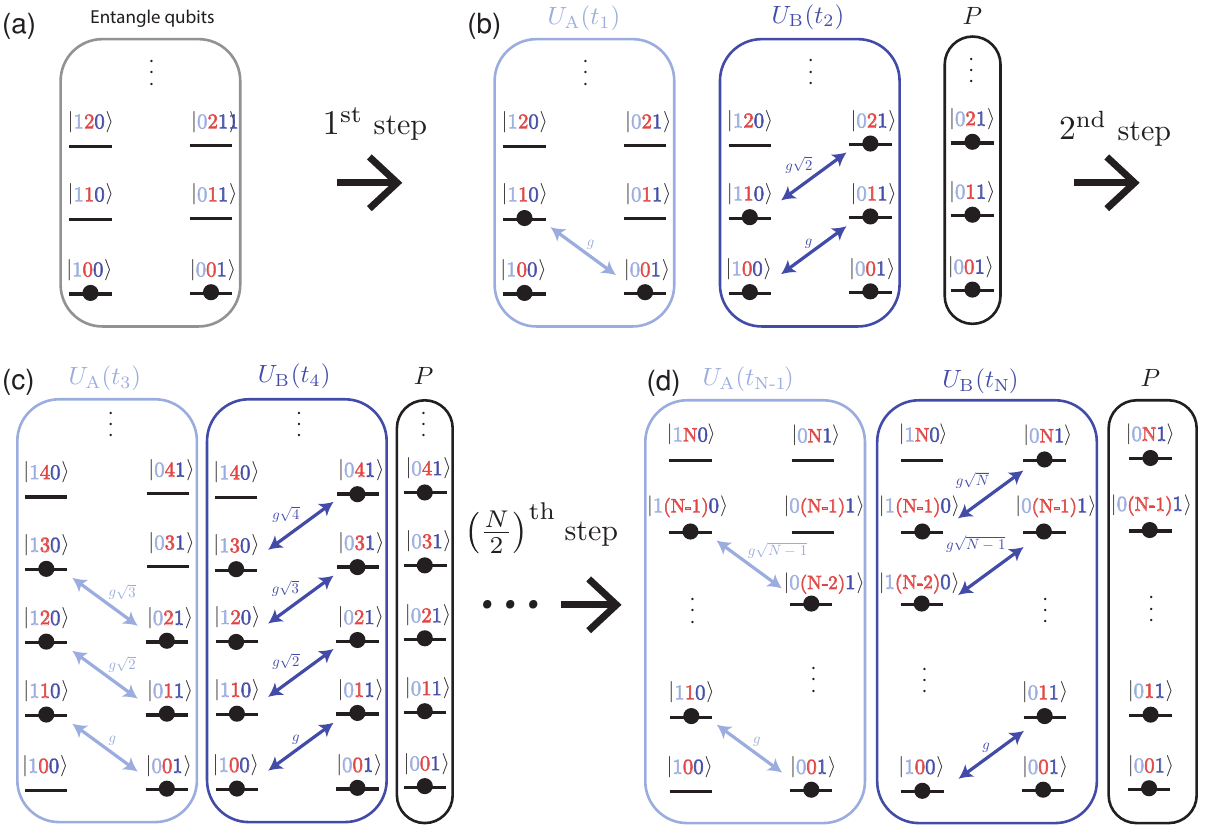}
  \caption{ {\bf Protocol for generating states with arbitrary phonon number probability distributions using flux pulsing and projective measurements.}
  (a) Assuming the mechanical resonator is in the ground state, we can start the protocol with the qubits in an entangled state. 
  This can be achieved by letting the qubits interact for a variable time without interacting with the resonator, e.g. by tuning the system such that the condition $J\geq|\omega_1-\omega_2|\gg\omega_\text{m}$ is satisfied.
  (b) The first step of the protocol is shown, which consists of three substeps:
  At first we turn on the interaction $\hat{H}_\text{A}=\hat{c}_1^\dagger\hat{b}^\dagger\hat{c}_2+\text{H.c.}$, by bringing the system in the resonance condition $\omega_2^+=\omega_1+\omega_\text{m}$, for variable time $t_1$ (light blue frame).
  Then, by tuning to $\omega_2^-=\omega_1-\omega_\text{m}$, the system evolves according to $\hat{H}_\text{B}=\hat{c}_1\hat{b}^\dagger\hat{c}_2^\dagger+\text{H.c.}$ for a different time $t_2$ (dark blue frame).
  Finally, we measure either qubit 1 or 2 and post-select on the outcome $|1_10_2\rangle$ (black frame).
  The freedom in choosing the coefficients of the initial two-qubit state together with the choice of interaction times $t_1$ and $t_2$ give enough degrees of freedom for generating a superposition state with arbitrary phonon number probability distribution up to the third level.
  (c) Second step of the protocol leading to a superposition state up to the fifth level.
  (d) General case for creating states with arbitrary phonon number probability distribution up to $N$, following $N/2$ steps.
 }
  \label{fig:ArbitraryStates}
\end{figure}

The protocol relies on initially preparing the two qubits in an arbitrary entangled state (Fig.\ref{fig:ArbitraryStates}a).
Assuming the resonator is in the ground state, the tripartite system is initially described by the following wavefunction
\begin{equation}
|\psi\rangle_0~=~(\alpha|0_11_2\rangle+\beta|1_10_2\rangle)|0_\text{m}\rangle,
\end{equation}
where $\alpha$, $\beta$ are complex numbers that can be chosen arbitrarily and are related by $|\alpha|^2+|\beta|^2=1$.
As described in the previous section, one can prepare this state by activating the exchange-type qubit-qubit interaction $U_\text{J}(t, \theta)$.
This can be done by detuning the qubits sufficiently such that their frequency difference is much greater than the mechanical frequency, and at the same time smaller than the direct qubit-qubit coupling, i.e. $J\geq|\omega_1-\omega_2|\gg\omega_\text{m}$, which can be adjusted by changing the flux $\Phi_\text{b}$ on the coupling SQUID.

The next step, schematically depicted in Fig.~\ref{fig:ArbitraryStates}b, consists of the following three substeps:\\
1. Flux pulse into resonance condition $\omega_2^+=\omega_1+\omega_\text{m}$ for variable time $t_\text{1}$\\
2. Flux pulse into resonance condition $\omega_2^-=\omega_1-\omega_\text{m}$ for variable time $t_\text{2}$,\\
3. Projective measurement on one of the qubits and post-selection on $|0_11_2\rangle$.\\
The initial wavefunction is transformed into
\begin{equation}
|\psi\rangle_1~=~\hat{P}\hat{U}_\text{B}(t_2)\hat{U}_\text{A}(t_1)|\psi\rangle_0,
\end{equation}
where $\hat{P}~=~|0_11_2\rangle\langle0_11_2|$ and $\hat{U}_\text{A,B}(t_{1,2})~=~e^{-i\hat{H}_\text{A,B}t_{1,2}}$, with interaction Hamiltonians $\hat{H}_\text{A}~=~\hat{c}_1^\dagger\hat{b}^\dagger\hat{c}_2+\text{H.c.}$ and $\hat{H}_\text{B}~=~\hat{c}_1\hat{b}^\dagger\hat{c}_2^\dagger+\text{H.c.}$, respectively.

Below we derive the resulting wavefunction after each substep:
\begin{align}
&1.~\hat{U}_\text{A}(t_1)|\psi\rangle_0~=~\alpha (\cos(gt_1)|0_10_\text{m}1_2\rangle-i\beta\sin(gt_1)|1_11_\text{m}0_2\rangle)+\beta|1_10_\text{m}0_2\rangle,\\
&2.~\hat{U}_\text{B}(t_2)\hat{U}_\text{A}(t_1)|\psi\rangle_0~=~\alpha \cos(gt_1)|0_10_\text{m}1_2\rangle+\beta\left(\cos(gt_2)|1_10_\text{m}0_2\rangle+\sin(gt_2)|0_11_\text{m}1_2\rangle\right)\nonumber\\
&\  \ \ \ \ \ \ \ \ \ \ \ \ \ \ \ \ \ \ \ \ \ \ \ \ \ \ \ \ \ \ \ \ \ +\alpha\sin(gt_1)\left(\cos(g\sqrt{2}t_2)|1_11_\text{m}0_2\rangle+\sin(g\sqrt{2}t_2)|0_12_\text{m}1_2\rangle\right)\\
&3.~\hat{P}\hat{U}_\text{B}(t_2)\hat{U}_\text{A}(t_1)|\psi\rangle_0~=~\frac{1}{P}\left(\alpha \cos(gt_1)|0_\text{m}\rangle+\beta\sin(gt_2)|1_\text{m}\rangle+\alpha\sin(gt_1)\sin(g\sqrt{2}t_2)|2_\text{m}\rangle\right)|0_11_2\rangle,
\end{align}
where $P$ is a normalisation factor determined by the Born rule.

Evidently, following this first step one has enough degrees of freedom ($\alpha, t_1, t_2$) to create states with arbitrary phonon number probability distribution up to three levels, $|\psi\rangle_1~=~\sum_{n=0}^2 c_n|n_\text{m}\rangle|0_11_2\rangle$.
Following a second step (Fig.~\ref{fig:ArbitraryStates}c), we find
\begin{align}
\hat{P}\hat{U}_\text{B}(t_4)\hat{U}_\text{A}(t_3)|\psi\rangle_1~=&~|0_11_2\rangle\{c_0\cos(gt_3)|0_\text{m}\rangle+c_1\cos(g\sqrt{2}t_3)\cos(gt_4)|1_\text{m}\rangle\nonumber\\
&\  \ \ \ +\left(c_0\sin(gt_3)\sin(g\sqrt{2}t_4)+c2\cos(g\sqrt{3}t_3)\cos(g\sqrt{2}t_4)\right)|2_\text{m}\rangle\nonumber\\
&\  \ \ \ +c_1\sin(g\sqrt{2}t_3)\sin(g\sqrt{3}t_4)|3_\text{m}\rangle+c_2\cos(g\sqrt{4}t_4)|4_\text{m}\rangle\}.
\end{align}
The addition of two more variables $t_3$, $t_4$ enables the creation of states with arbitrary phonon number probability distribution up to $|4\rangle$.

Suppose we start from an arbitrary mechanical state $|\psi\rangle~=~\sum_{n=0}^{N-2} c_n|n_\text{m}\rangle|0_11_2\rangle$, which can be created by applying the above protocol $(N/2-1)$ times.
Then, following another step we have 
\begin{align}
\prod_{j=N/2}^1\hat{P}\hat{U}_\text{B}(t_\text{2j})\hat{U}_\text{A}(t_\text{2j-1})|\psi\rangle_0~=~\hat{P}\hat{U}_\text{B}(t_\text{N-1})\hat{U}_\text{A}(t_\text{N})\left(\sum_{n=0}^{N-2} c_n|n\rangle|0_11_2\rangle\right)~=~\sum_{n=0}^{N} c_n^\prime|n\rangle|0_11_2\rangle,
\end{align}
where the new coefficients are determined by the previous ones according to the following relations:
\begin{align}
c_0^\prime~&=~c_0\cos(gt_\text{N-1})\\
c_1^\prime~&=~c_1\cos(g\sqrt{2}t_\text{N-1})\cos(gt_\text{N})\\
c_2^\prime~&=~c_0\sin(gt_\text{N-1})\sin(g\sqrt{2}t_\text{N})+c_2\cos(g\sqrt{3}t_\text{N-1})\cos(g\sqrt{2}t_\text{N})\\
&\ \ .\nonumber\\
&\ \ .\nonumber\\
&\ \ .\nonumber\\
c_n^\prime~&=~c_{n-2}\sin(g\sqrt{n-1}t_\text{N-1})\sin(g\sqrt{n}t_\text{N})+c_n\cos(g\sqrt{n+1}t_\text{N-1})\cos(g\sqrt{n}t_\text{N})\\
&\ \ .\nonumber\\
&\ \ .\nonumber\\
&\ \ .\nonumber\\
c_{N-1}^\prime~&=~c_{N-3}\sin(g\sqrt{N-2}t_\text{N-1})\sin(g\sqrt{N-1}t_\text{N})\\
c_{N}^\prime~&=~c_{N-2}\sin(g\sqrt{N-1}t_\text{N-1})\sin(g\sqrt{N}t_\text{N}).
\end{align}
Therefore, applying this protocol $N/2$ times can lead to the creation of states with arbitrary phonon number probability distribution up to $|N\rangle$.

\section{Validity for non-ideal system parameters}

\begin{figure}[h]
    \centering
  \includegraphics[width=0.5\linewidth]{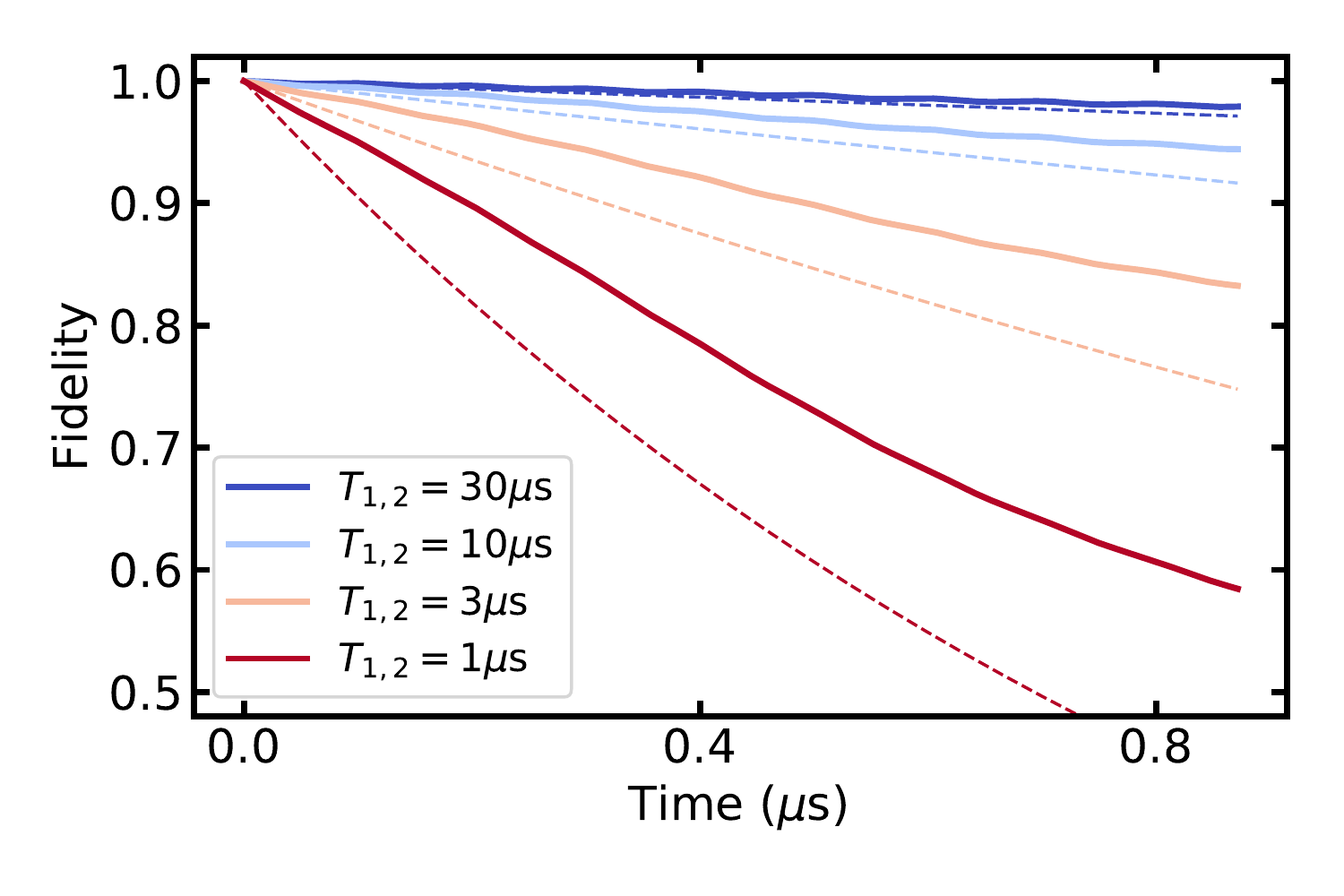}
  \caption{ {\bf Effect of qubit coherence on quantum state preparation.}
  Evolution of the fidelity of the prepared state to the ideal state, $\cos(gt)|0_10_\text{m}1_2\rangle -i \sin(gt)|1_11_\text{m}0_2\rangle$, using the protocol presented in Fig.~3, for different values of relaxation and pure dephasing times (assuming $T_1=T_2$).
  The dashed curves correspond to $e^{-t/T_1}$ for each case.
  }
  \label{fig:FidelVsT1}
\end{figure}

Although in the simulations we have considered realistic parameters taken from recent experiments, we realise that in an experimental scenario system parameters such as the qubit coherence may significantly deviate from the ones considered in Table~1.
Although bad qubit coherence may not affect the success of the cooling protocol, it can pose limits on the quantum state preparation protocols.
We have therefore examined the dependence of the fidelity of the prepared quantum state in Fig.~3 to the ideal evolution $|\psi_\text{ideal}\rangle=U(t)|0_10_\text{m}1_2\rangle= \cos(gt)|0_10_\text{m}1_2\rangle -i \sin(gt)|1_11_\text{m}0_2\rangle$ on the qubit coherence.
In Fig.~\ref{fig:FidelVsT1} we plot the evolution of the fidelity of the prepared density matrix $\rho$ to the ideal one $\rho_\text{ideal}=|\psi_\text{ideal}\rangle\langle\psi_\text{ideal}|$, for different values of $T_1$ (assuming $T_2=T_1$).
The dashed curves correspond to $e^{-t/T_1}$.
We find that high-fidelity state preparation ($>90~\%$) can be achieved with qubit coherence times of $T_1,~T_2\sim10~\mu$s which are standard in the superconducting qubit community and achievable in the presence of 10~mT magnetic fields~\cite{schneider2019transmon}.

\begin{figure}[h]
    \centering
  \includegraphics[width=1\linewidth]{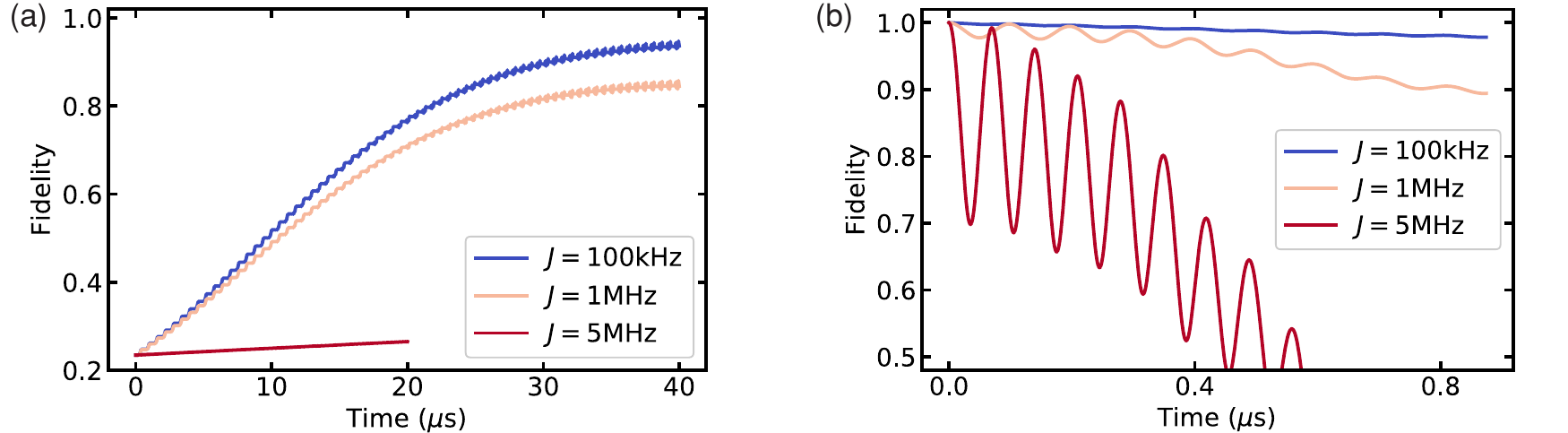}
  \caption{ {\bf Effect of flux fluctuations and stray qubit-qubit coupling on ground-state cooling and quantum state preparation.}
  (a) Evolution of the fidelity of the prepared mechanical state to the vacuum state using the cooling protocol of Fig.~2, for different values of qubit-qubit coupling $J~=~100$~kHz, $1$~MHz and $5$~MHz, corresponding to flux bias fluctuations of $\delta\Phi_\text{b}/\Phi_0~=~10^{-6},~10^{-5}$ and $5\times10^{-5}$, respectively.
  The same number of cooling steps were used in all simulations, where for $J=5$~MHz a cooling step of 100~ns instead of 200~ns was chosen as an optimal cooling time.
  (b) Evolution of the fidelity of the prepared state to the ideal state, $\cos(gt)|0_10_\text{m}1_2\rangle -i \sin(gt)|1_11_\text{m}0_2\rangle$, following the protocol used in Fig.~3, for the same variations on $J$ and $\delta\Phi_\text{b}/\Phi_0$ as in (a).
  }
  \label{fig:FidelVsJ}
\end{figure}

One effect that is not considered in the main text is that of fluctuations on the flux bias channel $\delta\Phi_\text{b}$.
These can occur as a result of environmental magnetic field noise or noise of the current source used for biasing the flux line.
Given the stability of our current sources and assuming flux noise levels reported in similar devices~\cite{kumar2016origin, hutchings2017tunable}, we estimate such fluctuations to be around $\delta\Phi_\text{b}/\Phi_0\sim10^{-6}$ during the course of the preparation protocols.
The most important effect of flux fluctuations would be to introduce a stray qubit-qubit exchange-type coupling as $J_\text{L}\propto E_{\text{J},\Sigma}^\text{c} \cos(\pi(\Phi_\text{b}+\delta\Phi_\text{b})/\Phi_0)$.
More specifically, adding a fluctuation of $\delta\Phi_\text{b}/\Phi_0\sim10^{-6}$ would result in $J_\text{L}\rightarrow J_\text{L}+100$~kHz, while $\delta\Phi_\text{b}/\Phi_0\sim10^{-5}$ and $\delta\Phi_\text{b}/\Phi_0\sim10^{-4}$ translate to additional coupling of 1 and 10~MHz, respectively.
The latter would be detrimental for the state preparation protocols as it is of the order of the mechanical frequency.
Note that the effect of flux noise on the qubit-qubit coupling is also amplified by our choice of a large coupling Josephson energy amplitude $E_{\text{J},\Sigma}^\text{c}/h=200$~GHz.
In Fig.~\ref{fig:FidelVsJ} we plot the effect of a finite qubit-qubit coupling on the cooling and quantum state preparation protocols.
We find that for $J<1$~MHz (therefore $\delta\Phi_\text{b}<10~\mu\Phi_0$) it is possible to maintain high-fidelity quantum state preparation.
Note that despite the sharp dependence of the tripartite coupling on the flux bias (Fig.~1), this change is less than $0.1~\%$ in the case of $\delta\Phi_\text{b}<10~\mu\Phi_0$.

\begin{figure}[h]
    \centering
  \includegraphics[width=0.5\linewidth]{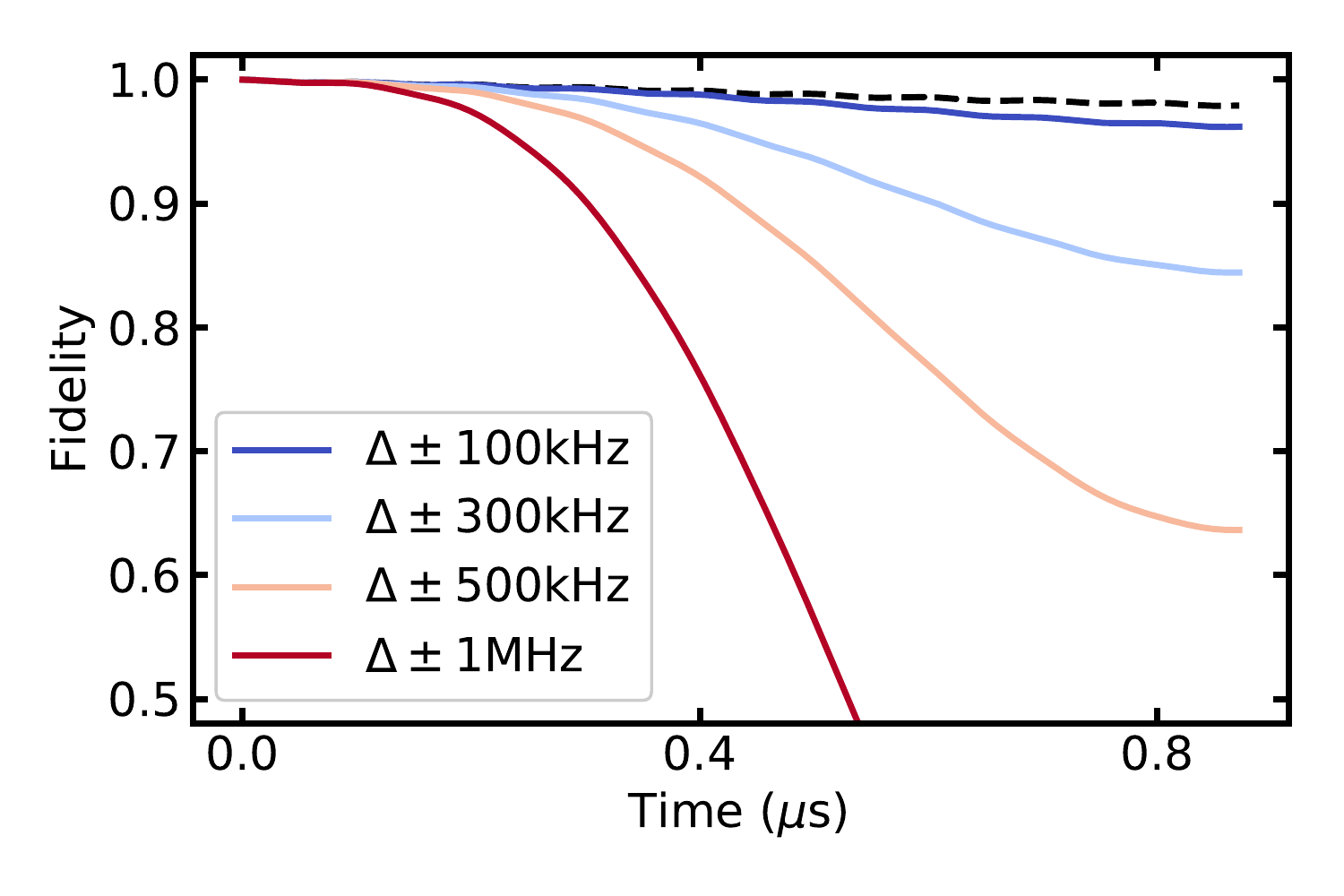}
  \caption{ {\bf Effect of imperfect flux pulsing on quantum state preparation.}
  Evolution of the fidelity of the prepared state to the ideal state using the protocol presented in Fig.3, for different variations of the qubit-qubit detuning $\Delta$.
  The black dashed curve corresponds to no variation, i.e. $\Delta=\omega_\text{m}$.
  }
  \label{fig:FidelVsFluxImperfections}
\end{figure}

Additionally, in Fig.~\ref{fig:FidelVsFluxImperfections} we examine the effect of imperfect qubit flux pulsing on the fidelity of the prepared quantum state.
We find that the fidelity of the protocol is sensitive to this parameter and that targeting the qubit frequencies within $\sim100$~kHz is required for high-fidelity quantum state preparation.

\end{document}